\documentclass[openacc]{rsproca_new}

\usepackage{booktabs}
\usepackage{xcolor}
\usepackage{float}

\newcommand{\FracInt}[5][\empty]{{\ifx#1\empty {_{#2}} \, \mathcal{I}\else{_{#2}^{#1}\mathcal{I}}\fi} \, {{}^{#4}_{#3}} #5 }
\newcommand{\FracDer}[5][\empty]{{\ifx#1\empty {_{#2}} \, \mathcal{D}\else{_{#2}^{#1}\mathcal{I}}\fi} \, {{}^{#4}_{#3}}  #5 }

\titlehead{Research}

\begin{document}

\title{Advanced creep modeling for polymers: A variable-order fractional calculus approach}

\author{José Geraldo Telles Ribeiro$^{1}$,\\ Americo Cunha Jr$^{2,3}$}

\address{$^{1}$ Department of Mechanical Engineering, Rio de Janeiro State University, Brazil\\
$^{2}$ Department of Applied Mathematics, Rio de Janeiro State University, Brazil\\
$^{3}$ National Laboratory of Scientific Computing, Brazil
}

\subject{mechanics of materials,
mathematical modelling,
mechanical engineering,
chemical engineering,
applied mathematics}


\corres{José Geraldo Telles Ribeiro\\
\email{jose.ribeiro@uerj.br}}

\begin{abstract}
Polymer-based plastics exhibit time-dependent deformation under constant stress, known as creep, which can lead to rupture or static fatigue. A common misconception is that materials under tolerable static loads remain unaffected over time. Accurate long-term deformation predictions require experimental creep data, but conventional models based on simple rheological elements like springs and dampers often fall short, lacking the flexibility to capture the power-law behavior intrinsic to creep processes. The springpot, a fractional calculus-based element, has been used to provide a power-law relationship; however, its fixed-order nature limits its accuracy, particularly when the deformation rate evolves over time. This paper introduces a variable-order springpot model that dynamically adapts to the evolving viscoelastic properties of polymeric materials during creep, capturing changes between glassy, transition, and rubbery phases. Model parameters are calibrated using a robust procedure for model identification based on the cross-entropy method, resulting in physically consistent and accurate predictions. This advanced modeling framework not only overcomes the limitations of fixed-order models but also establishes a foundation for applying variable-order mechanics to other viscoelastic materials, providing a valuable tool for predicting long-term material performance in structural applications.
\end{abstract}

\maketitle

\section{Introduction}

\begin{figure}[h!]
\centering
\includegraphics[width=120mm]{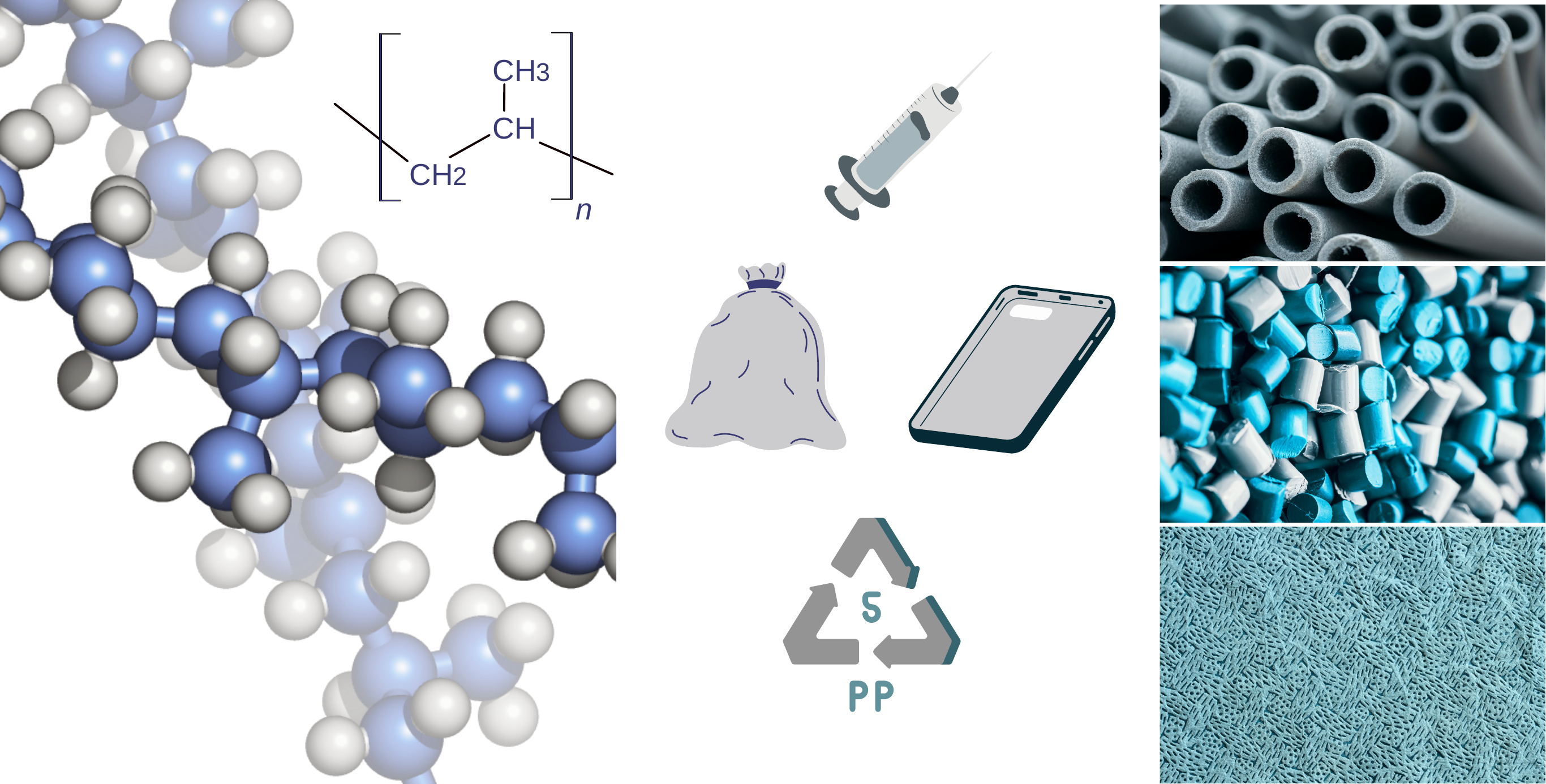}
\caption{Molecular structure of a polymer alongside common products and structures made from polymeric materials. These materials are widely used in various applications due to their exceptional mechanical properties and resistance to stress and fatigue. They are also recyclable, adding to their environmental significance in industrial applications.}
\label{Fig01}
\end{figure}

Polymeric materials (Fig.~\ref{Fig01}) are extensively utilized across various industrial and commercial sectors due to their distinctive combination of mechanical properties, chemical resistance, and cost-effectiveness \cite{crawford2020plastics,Dowling2012}. Their low density and high tensile strength make them particularly suitable for applications in the automotive industry, such as bumpers, interior components, and battery cases, where lightweight materials are essential for enhancing fuel efficiency. Their excellent resistance to moisture, chemicals, and fatigue also makes them a preferred choice for packaging, piping, and medical devices. Additionally, their thermal stability and ability to endure repeated cycles of stress and strain are advantageous in applications where creep behavior, as analyzed in this study, is critical. The long molecular chains of polypropylene (PP) or polyvinyl chloride (PVC) contribute to their mechanical complexity, making the deformation and creep behavior of objects manufactured from these materials highly nonlinear and time-dependent. The versatility of these polymers is further enhanced by their ease of processing, recyclability, and low production costs, making them attractive materials for manufacturers across various industries.

Creep, the time-dependent deformation under prolonged stress, is a key concern for PP and PVC, as it can compromise structural integrity over time \cite{crawford2020plastics}. This deformation is especially problematic in load-bearing components or applications exposed to constant forces, such as piping systems or storage containers. Prolonged exposure to stress can lead to mechanical failure, dimensional instability, and functional degradation, especially in high-temperature environments where creep effects are amplified \cite{Dowling2012,castro2016fatigue}. Understanding and accurately predicting creep behavior is therefore essential to ensure the long-term reliability and safety of polymeric components.

\begin{figure}[h!]
\centering
\includegraphics[width=120mm]{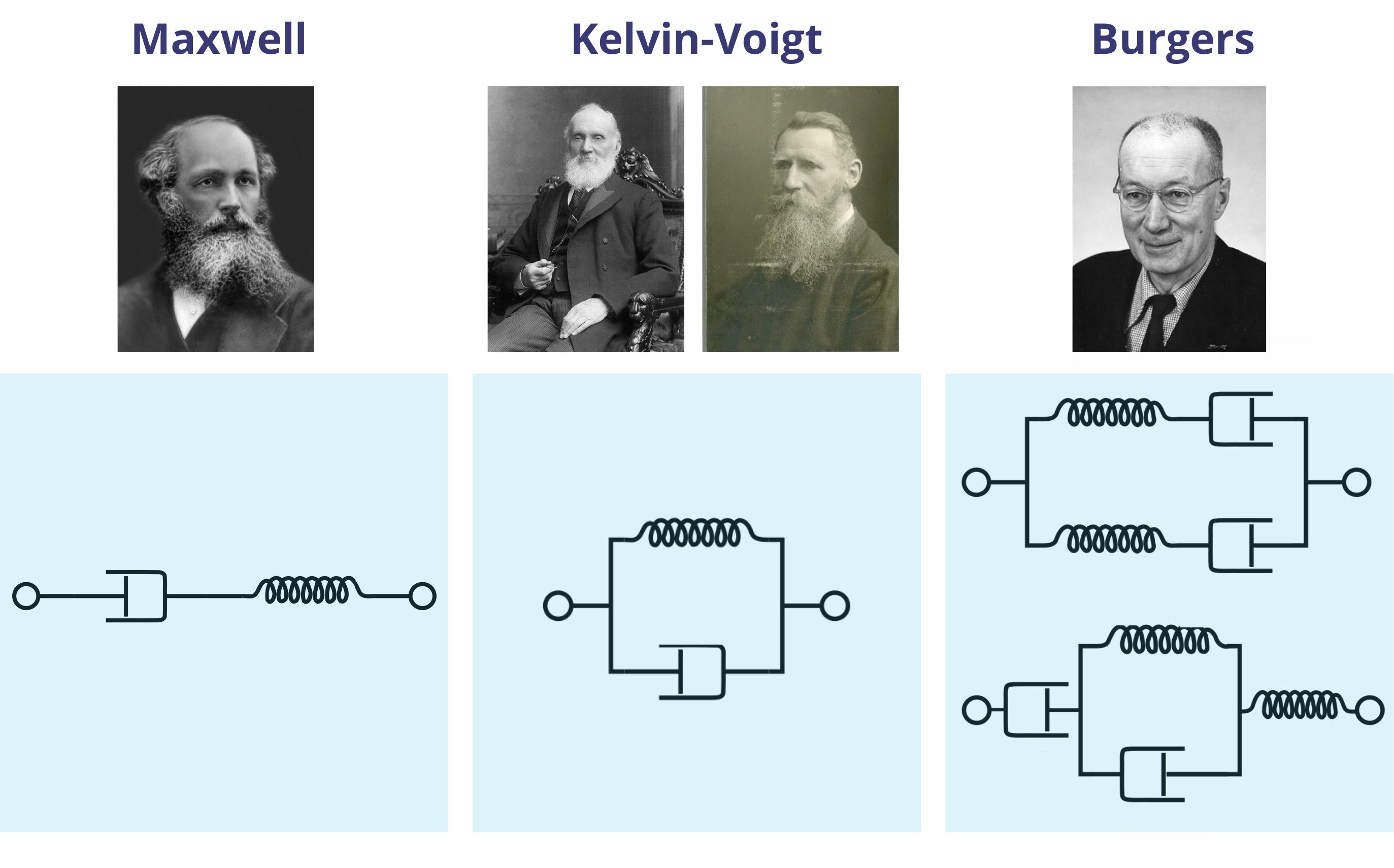}
\caption{Schematic representations of some classic constitutive models for viscoelastic materials like polymers, including the Maxwell, Kelvin-Voigt, and Burgers models, illustrating the combination of spring and dashpot elements used to describe the material's creep behavior.}
\label{Fig02}
\end{figure}

Traditionally, creep has been modeled using combinations of springs and dashpots, as represented in Fig.~\ref{Fig02} \cite{morro2017modelling,Lakes2010}. Basic models like the Maxwell model (elements in series) and the Kelvin-Voigt model (elements in parallel) use only a spring and dashpot, making them limited in their ability to replicate the complex time-dependent deformation observed experimentally. To address these limitations, more sophisticated models like the generalized Maxwell and Burgers models have been developed, incorporating additional components for improved fit to experimental data \cite{di2011visco}. However, these classical models often reduce creep behavior to sums of exponential terms, failing to capture the nonlinear and power-law characteristics of polymer deformation observed in real-world conditions.

\begin{figure}[h!]
\centering
\includegraphics[width=75mm]{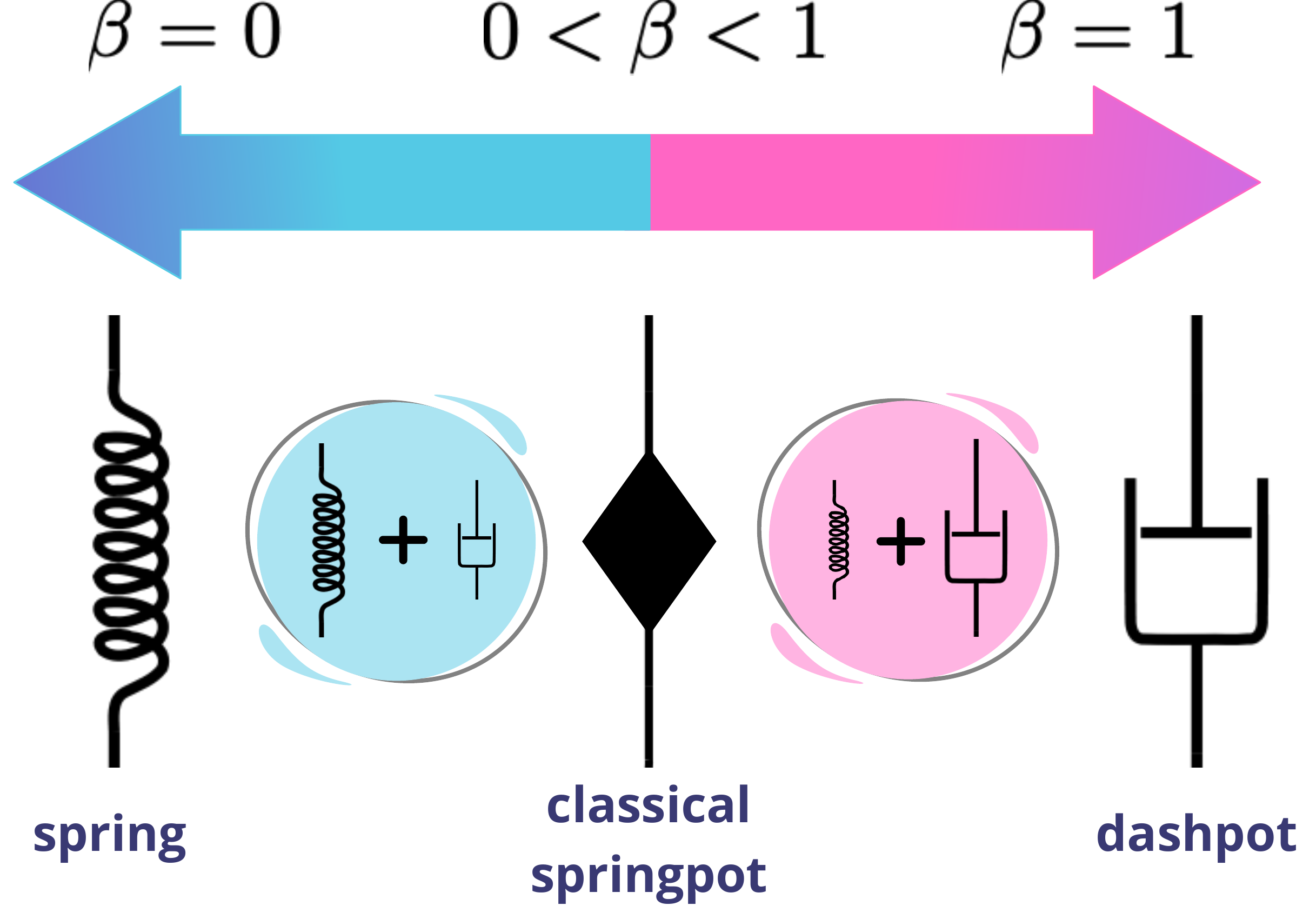}
\caption{ Schematic representation of the springpot model, illustrating its intermediate behavior between a spring (purely elastic) and a dashpot (purely viscous). The springpot is used to model viscoelastic behavior, where deformation is both time-dependent and elastic.}
\label{Fig03}
\end{figure}

Early experiments on polymers demonstrated that time-dependent behavior is better described by power-law functions, which traditional rheological models cannot represent accurately \cite{nutting1921new, gemant1936method}. To overcome this, fractional calculus has been introduced as a more flexible framework for modeling viscoelastic behavior. The fractional calculus approach led to the creation of the \emph{springpot} model, an element that interpolates between purely elastic and purely viscous behavior using a fractional order parameter $0 \leq \beta \leq 1$ (Fig.~\ref{Fig03}) \cite{mainardi2008time}. The springpot model, with its parameters $\beta$ and generalized viscosity $C_\beta > 0$, provides a more effective representation of viscoelastic materials by capturing intermediate behaviors that standard models cannot \cite{blair1944analytical, ortigueira2015fractional}.

\pagebreak
Initial applications of the springpot involved replacing dashpots in classical models \cite{morro2017modelling, di2011visco}, leading to fractional models with fewer parameters, such as the fractional Kelvin-Voigt and fractional Maxwell models. These fractional models demonstrated improved performance in specific scenarios; for example, in \cite{ribeiro2021modeling}, the fractional Kelvin-Voigt model was used to simulate the creep behavior of polypropylene. However, it struggled to accurately capture the initial creep response immediately after load application. To address these limitations, more recent studies have explored combinations of multiple springpots to enhance model accuracy across various time scales \cite{bonfanti2020fractional, jozwiak2015fractional}.

A key assumption in many of these models is that material properties remain constant during the creep process. In practice, however, polymers may undergo changes in mechanical properties due to microstructural modifications, such as chain stretching, orientation, and strengthening at the molecular level \cite{di2020novel}. One particularly important aspect is the time-dependence of viscosity. In this context, \cite{su2020fractional} analyzed the physical significance of fractional models, proposing an equivalence between the fractional Maxwell model and the classical Maxwell model with variable viscosity. This analysis was further extended to the Kelvin-Voigt model in \cite{gao2023bridge}.

Many physical phenomena that are effectively modeled using fractional calculus benefit from the use of a variable-order (time-dependent) approach, rather than a fixed-order framework \cite{garrappa2021variable}. The need for variable-order fractional calculus in accurately capturing complex time-dependent behaviors has been discussed in various studies, including a comprehensive review of related phenomena in \cite{patnaik2020applications}. This concept, first introduced in \cite{samko1993integration} and further developed in \cite{samko1995fractional, samko2013fractional}, enables a flexible representation of dynamic systems.

In the field of viscoelastic materials, variable-order fractional calculus has been applied to model evolving mechanical behavior in several contexts. Examples include viscoelastic properties of composites \cite{ramirez2007variable}, sintered nano-silver paste under tensile and shear stress \cite{cai2020variable}, rock creep \cite{gao2021full}, and polymer viscoelasticity \cite{wang2023fractional}. Additionally, this approach has been utilized to study damping materials \cite{li2022variable}, the behavior of rubbery and glassy polymers \cite{meng2019variable, meng2020parameter}, natural fiber polymer composites \cite{xiang2020creep}, edge dislocations \cite{patnaik2020variable}, fracture mechanics \cite{patnaik2021variable}, and the aging of concrete \cite{beltempo2018fractional}. These studies underscore the effectiveness of variable-order calculus in providing a detailed, adaptable framework for modeling time-dependent mechanical behavior.

In this study, a new type of fractional calculus-based model is proposed for the springpot rheological element, this time considering a time-dependent order $\beta$ during the creep process. This variation induces a simultaneous change in the material parameter $C_\beta$, while the viscosity $\eta > 0$ and elastic modulus $E > 0$ remain constant throughout the process. Creep curves of polypropylene and polyvinyl chloride at 20°C and under various load levels are used as reference data for validating the model. Ultimately, a set of equations is derived to predict the deformation of PP and PVC as a function of applied load and time, providing an accurate tool for calculating time-dependent behavior under different stress conditions.

\section{Fractional Calculus}

\subsection{Constant-order operators}

Fractional calculus extends classical calculus by allowing integrals and derivatives to take non-integer (fractional) orders, making it highly versatile for modeling viscoelastic and other complex phenomena \cite{machado2011recent,machado2014development,delia2021unified}. One widely used formulation is the Riemann-Liouville fractional integral, chosen here for its ability to generalize integer-order integrals to fractional orders, which provides a balance between computational efficiency and model accuracy.

The Riemann-Liouville fractional integral can be derived from Cauchy's formula
\begin{equation}
\int_{0}^{t} \, \int_{0}^{\tau_{n-1}} \cdots \int_{0}^{\tau_1} f(\tau) \, d\tau \, d\tau_1 \, \cdots \, d\tau_{n-1} = \displaystyle \frac{1}{(n-1)!} \, \int_{0}^{t} (t-\tau)^{n-1} \, f(\tau) \, d\tau \, .
\label{eq_Cauchy}
\end{equation}

By generalizing the integer $n \in \mathbb{Z}$ to a fractional real value $\alpha \in \mathbb{R}$, we obtain the Riemann-Liouville fractional integral
\begin{equation}
\FracInt{}{}{\alpha}{f(t)} = \displaystyle \frac{1}{\Gamma(\alpha)} \, \int_{0}^{t} (t-\tau)^{\alpha-1} \, f(\tau) \, d\tau \quad \text{for} \quad t > 0 \, ,
\label{eq_Riemann_integra}
\end{equation}
where $\Gamma(\alpha)$ denotes the gamma function, a generalization of factorials for non-integer values, giving this operator its distinctive flexibility.

The Riemann-Liouville fractional derivative of order $\alpha$ is obtained by applying a fractional integral of order $n - \alpha$, followed by $n$ integer differentiations
\begin{equation}
\FracDer{}{}{\alpha}{f(t)} = \displaystyle \frac{1}{\Gamma(n - \alpha)} \, \frac{d^n}{dt^n} \left( \int_{0}^{t} \displaystyle \frac{f(\tau)}{(t-\tau)^{\alpha-n+1}} \, d\tau \right) \quad \text{for} \quad  t > 0 \quad \text{and} \quad n-1 \leq \alpha \leq n \, .
\label{eq_Riemann_derivative}
\end{equation}

A more generalized approach involves fractional integration with respect to another function $g(t)$, allowing fractional orders to apply to non-constant weight functions, giving us
\begin{equation}
\FracInt[g]{}{}{\alpha}{f(t)} = \frac{1}{\Gamma(\alpha)} \, \int_{0}^{t} g'(\tau) \, \left(g(t)-g(\tau)\right)^{\alpha-1} \, f(\tau) \, d\tau \quad \text{for} \quad  t > 0 \, .
\label{eq_fractional_another}
\end{equation}

When $g(t) = t$, this expression reduces to the Riemann-Liouville fractional integral, and for $g(t) = \ln t$, it becomes the Hadamard fractional integral \cite{colombaro2018scott}
\begin{equation}
\FracInt[H]{}{}{\alpha}{f(t)} = \frac{1}{\Gamma(\alpha)} \, \int_{0}^{t} \frac{1}{\tau} \, \left(\ln{\frac{t}{\tau}} \right)^{\alpha-1} f(\tau) \, d\tau \quad \text{for} \quad  t > 0 \, .
\label{eq_fracational_Hadamard}
\end{equation}

These extensions illustrate the adaptability of fractional calculus in modeling non-local processes, accommodating complex temporal dependencies often seen in viscoelastic materials.

\subsection{Variable-order operators}

Variable-order (VO) fractional calculus is an extension of fractional calculus where the integral or derivative order varies with time or other variables, enabling the model to adapt dynamically based on system conditions \cite{patnaik2020applications}. This flexibility allows variable-order fractional integrals to better model complex, nonlinear, and time-varying processes observed across many fields, such as viscoelasticity, control systems, and biological processes \cite{garrappa2021variable, patnaik2020applications}.

The VO fractional integral of a function $f(t)$ is defined similarly to the constant-order integral, with a variable order $\alpha(t) \in \mathbb{R}$ \cite{samko1993integration,samko1995fractional,samko2013fractional}:
\begin{equation}
\FracInt{}{}{\alpha(t)}{f(t)} = \frac{1}{\Gamma[\alpha(t)]} \, \int_{0}^{t} (t-\tau)^{\alpha(t)-1} \, f(\tau) \, d\tau \quad \text{for} \quad t > 0 \, .
\label{eq_Riemann_VO}
\end{equation}

Similarly, fractional integration of $f(t)$ with respect to another function $g(t)$ can be generalized to a variable order as
\begin{equation}
\FracInt[g]{}{}{\alpha(t)}{f(t)} = \frac{1}{\Gamma[\alpha(t)]} \, \int_{0}^{t} 
g'(\tau) \, \left(g(t)-g(\tau)\right)^{\alpha(t)-1} f(\tau) \, d\tau \quad \text{for} \quad t > 0 \, .
\label{eq_frac_f_g_VO}
\end{equation}

For $g(t) = \ln t$, we obtain the variable-order Hadamard operator \cite{almeida2015computing}
\begin{equation}
\FracInt[H]{}{}{\alpha(t)}{f(t)} = \frac{1}{\Gamma\left[ \alpha(t)  \right]} \, \int_{0}^{t} \frac{1}{\tau} \, \left(\ln{\frac{t}{\tau}} \right)^{\alpha(t)-1} f(\tau) \, d\tau \quad \text{for} \quad  t > 0 \, .
\label{eq_fracational_Hadamard_VO}
\end{equation}

Variable-order calculus is valuable mathematical tool in diverse fields like materials science, where it models time-evolving material properties, and biology, where reaction rates change dynamically \cite{garrappa2021variable, patnaik2020applications}. By adjusting the order based on system conditions, it effectively captures complex, time-dependent behaviors.

\pagebreak
\section{Fractional Viscoelastic Rheological Models}
\label{sec3}

\subsection{The constant-order springpot}

Boltzmann's Superposition Principle is a fundamental concept in viscoelasticity used to describe the time-dependent behavior of materials under load. This principle is essential when analyzing materials that exhibit both elastic (instantaneous) and viscous (time-dependent) deformation, such as polymers, biological tissues, and certain metals at elevated temperatures. In a simple uniaxial stress scenario, the strain response $\varepsilon(t)$ due to a stress history $\sigma(t)$ is described by the convolution integral
\begin{equation}
\varepsilon (t)=\int_{0}^{t} J(t-\tau) \, \frac{d\sigma}{d\tau}(\tau) \, d\tau 
\label{eq_Bolt_def}
\end{equation}
where $\varepsilon(t)$ is the strain at time $t$; $\sigma(\tau)$ is the applied stress at time $\tau$; and $J(t - \tau)$ is the \textit{creep compliance function}, describing the material's time-dependent response to stress applied at $\tau$.

Experimental results from the early 20th century \cite{nutting1921new, gemant1936method} demonstrated that the creep behavior of viscoelastic materials can be modeled using power-law functions:
\begin{equation}
\varepsilon \left( t \right) = \frac{t^\beta}{C_\beta \, \Gamma(\beta +1)} \, \sigma_0 \, ,  \quad  0 \leq \beta \leq 1 \, ,
\label{eq_def_creep}
\end{equation}
where the stress $\sigma_0 > 0$ is an applied load. For this case, the creep compliance function is
\begin{equation}
J \left( t \right) = \frac{t^\beta}{C_\beta \, \Gamma(\beta +1)} \, , \quad 0 \leq \beta \leq 1 \, .
\label{eq_J_creep}
\end{equation}

This can be seen by substituting Eq.~(\ref{eq_J_creep}) into Eq.~(\ref{eq_Bolt_def}), where we get
\begin{equation}
\varepsilon(t)=\frac{1}{C_\beta \, \Gamma(\beta +1)} \, \int_{0}^{t} (t-\tau)^\beta \, \frac{d\sigma}{d\tau}(\tau) \, d\tau \, .
\label{eq_pot_Bolt_01}
\end{equation}

If the change of variables $\beta = \alpha - 1$ is made in Eq.~(\ref{eq_pot_Bolt_01}), it is obtained
\begin{equation}
\varepsilon(t) = \frac{1}{C_\beta \, \Gamma(\alpha)} \, \int_{0}^{t} (t-\tau)^{\alpha-1 } \, \frac{d\sigma}{d\tau}(\tau) \, d\tau
\label{eq_pot_Bolt_02}
\end{equation}
which aligns with the Riemann-Liouville fractional integral definition shown in Eq.~(\ref{eq_Riemann_integra}). Thus, the constitutive equation of the springpot relating $\varepsilon(t)$ and $\sigma(t)$  can be written as
\begin{equation}
\varepsilon(t) = \frac{1}{C_\beta} \, \FracInt{}{}{\alpha}{\frac{d\sigma}{dt}(t)} = \frac{1}{C_\beta} \, \FracInt{}{}{\beta+1}{\frac{d\sigma}{dt}(t)} = \frac{1}{C_\beta} \, \FracInt{}{}{\beta}{\sigma(t)} \, ,
\label{eq_pot_integral_01}
\end{equation}
and then
\begin{equation}
\varepsilon(t)= \frac{1}{C_\beta \, \Gamma(\beta)} \, \int_{0}^{t} (t-\tau)^{\beta-1} \, \sigma(\tau) \, d\tau \, .
\label{eq_pot_integral_02}
\end{equation}

Taking the fractional derivative of order $\beta$ on both sides of Eq.~(\ref{eq_pot_integral_01}), and recalling that $\FracDer{}{}{\beta}{\FracInt{}{}{\beta}{\sigma(t)}} = \sigma(t)$, the constitutive equation of the springpot can also be expressed as
\begin{equation}
\sigma (t) = C_\beta \, \FracDer{}{}{\beta}{\varepsilon(t)} \, ,  \ \ \ \ 0\leq \beta \leq 1 \, .
\label{eq_pot_derivative}
\end{equation}

For a constant stress history $\sigma(t) = \sigma_0$ the solution of Eq.(\ref{eq_pot_integral_02}) is given by (\ref{eq_def_creep}), showing that the springpot constitutive model is in agreement with the experimental evidence \cite{nutting1921new, gemant1936method}.

The parameters $C_\beta$ and $\beta$ are intrinsic properties that influence the material’s response under stress. For $\beta = 0$, the element behaves elastically, with $C_\beta$ equivalent to the elastic modulus $E$ (i.e., $C_{\beta=0} = E$), while for $\beta = 1$, the element exhibits purely viscous behavior with $C_\beta$ representing the viscosity $\eta$ (i.e., $C_{\beta=1} = \eta$).

\pagebreak
Additionally, $C_\beta$ represents a combination of dissipative and elastic elements in the material, with $\beta$ indicating their relative predominance. It is noteworthy that $C_\beta$ has units of $\text{Pa} \cdot \text{s}^\beta$ and can be referred to as the \emph{firmness} of the material \cite{bonfanti2020fractional}:
\begin{equation}
C_\beta = \eta^{\beta} \, E^{1-\beta} = T^\beta \, E \, ,
\label{eq_Cbeta}
\end{equation}
where $T=\eta/E$ is the \emph{material timescale}. Substituting Eq.(\ref{eq_Cbeta}) in Eq.(\ref{eq_pot_integral_02}) leads to
\begin{equation}
\varepsilon(t) = \frac{1}{E \, \Gamma(\beta)} \int_{0}^{t} \frac{1}{T} \, \left( \frac{t}{T}-\frac{\tau}{T} \right)^{\beta-1 } \, \sigma(\tau) \, d\tau \, ,
\label{eq_pot_integral_2}
\end{equation}
that is the related to fractional integration of $\sigma(t)$ with respect to the function $g(t)=t/T$, i.e., 
\begin{equation}
\varepsilon(t) = \frac{1}{E} \, \FracInt[g]{}{}{\beta}{\sigma(t)} \, .
\label{eq_pot_integral_3}
\end{equation}

To derive the constant-order springpot model parameters $\beta$ and $C_\beta$ from experimental data we perform a linear fit on the logarithmic form of Eq.~(\ref{eq_def_creep}):
\begin{equation}
\log\frac{\varepsilon(t)}{\sigma_0} = \beta \log t - \log \left( C_\beta \Gamma(\beta+1) \right) \, .
\label{eq_def_creep_log}
\end{equation}

\begin{figure}[h!]
\centering
\includegraphics[width=130mm]{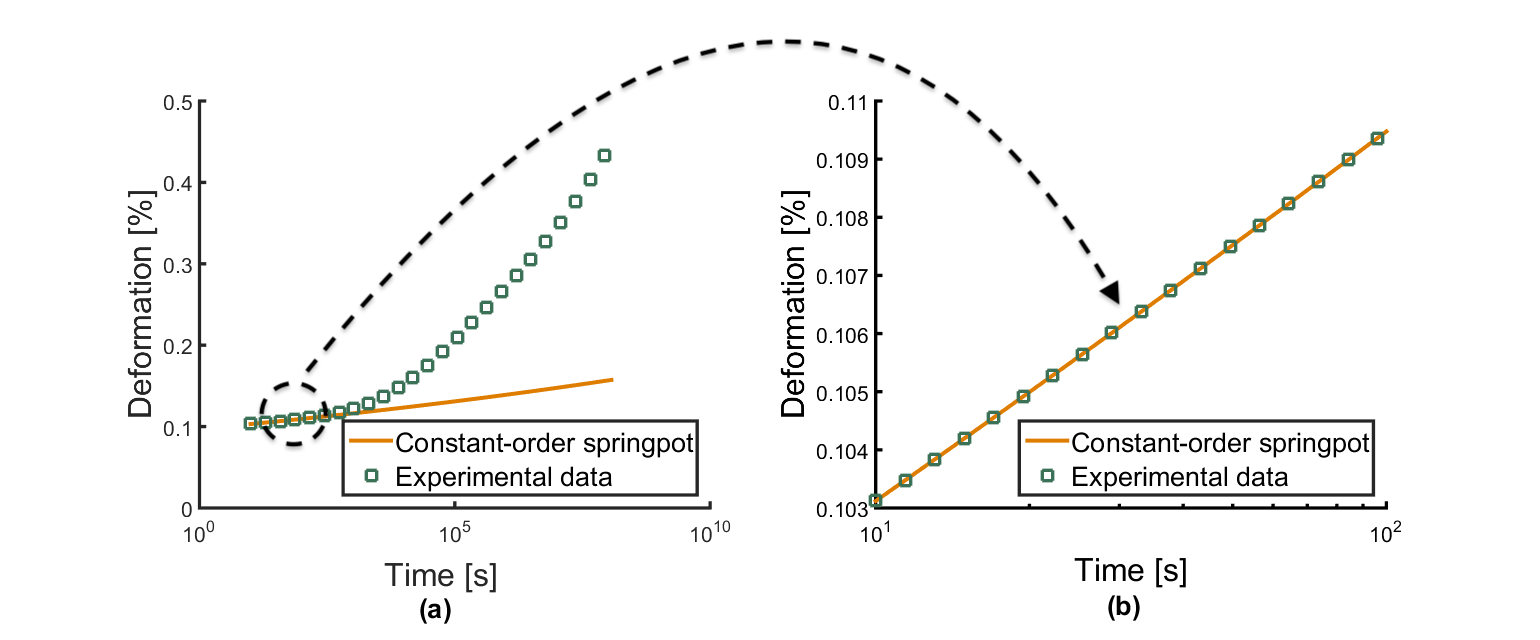}
\caption{The creep curve of PP at a temperature of $20\text{ºC}$ under a load of $\sigma_0 = 1.4 \text{ MPa}$ is compared to the constant-order springpot model. The parameters were estimated based on the time interval between 10 and 140 seconds, yielding $\beta=0.0260$ and $C_\beta$ = 1460 $\text{MPa} \cdot \text{s}^\beta$. In (a), it is evident that the model does not provide a good fit for long-term creep behavior. However, (b) shows that the model fits well within the specified interval.}
\label{Fig04}
\end{figure}

\begin{figure}[h!]
\centering
\includegraphics[width=130mm]{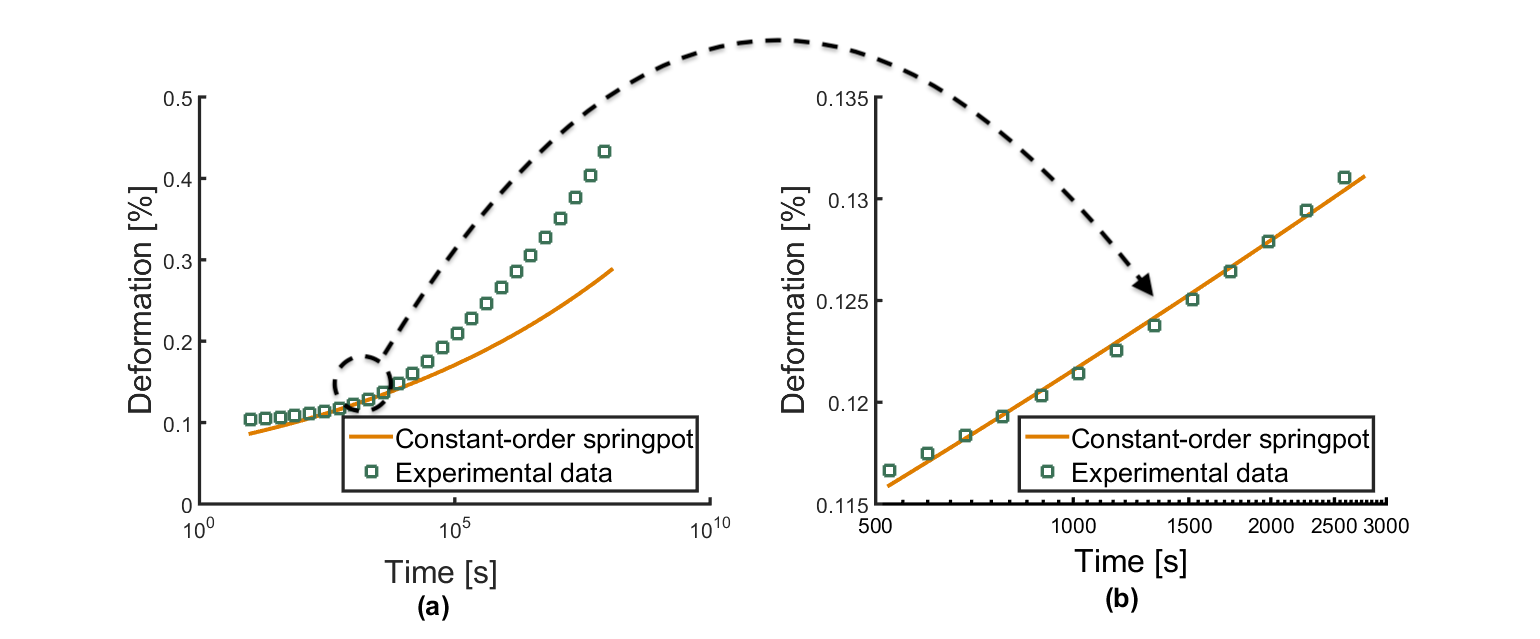}
\caption{The creep curve of PP at a temperature of $20\text{ºC}$ under a load of $\sigma_0 = 1.4 \text{ MPa}$ is compared to the constant-order springpot model. The parameters were estimated based on the time interval between 8 and 46 min, yielding $\beta = 0.0738$ and $C_\beta$ = 2000 $\text{MPa} \cdot \text{s}^\beta$. In (a), it is evident that the model does not provide a good fit for long-term creep behavior. However, (b) shows that the model fits well within the specified interval.}
\label{Fig05}
\end{figure}

\begin{figure}[h!]
\centering
\includegraphics[width=130mm]{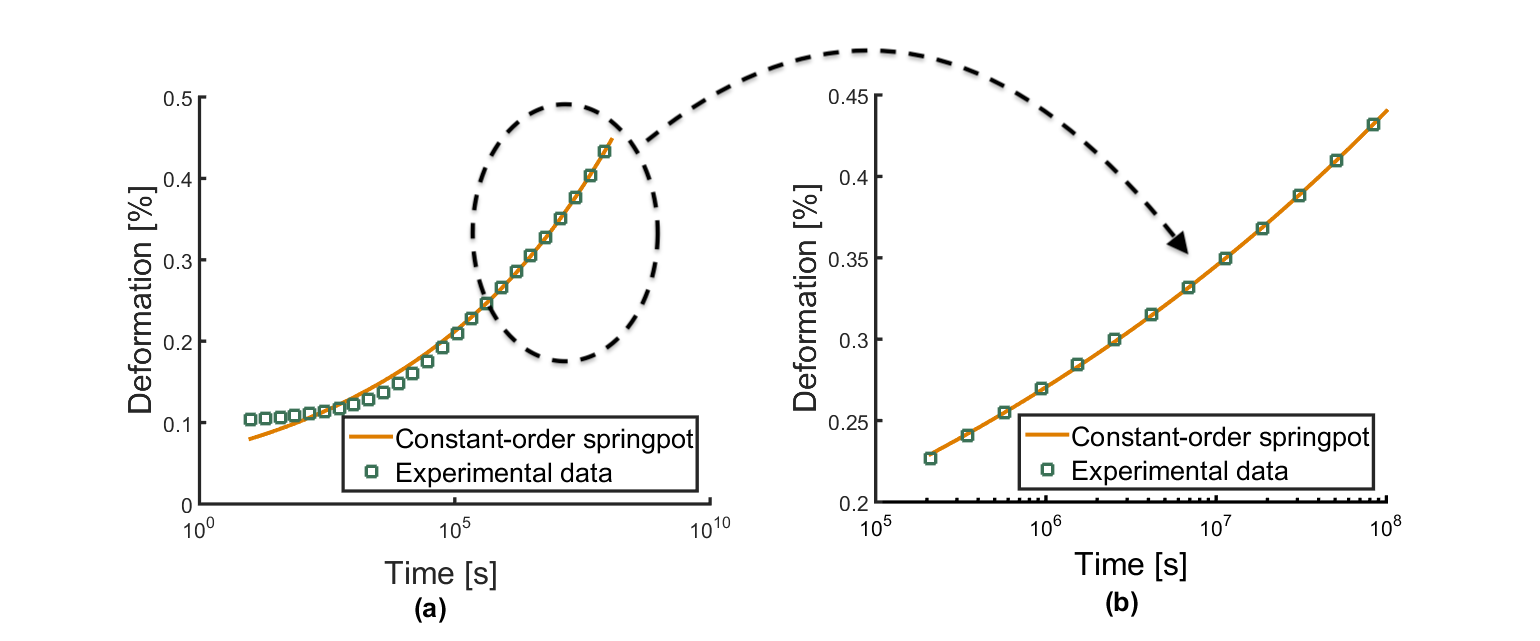}
\caption{The creep curve of PP at a temperature of $20\text{ºC}$ under a load of $\sigma_0 = 1.4 \text{ MPa}$ is compared to the constant-order springpot model. The parameters were estimated based on the time interval between 0 and 58 hours, yielding $\beta=0.1057$ and $C_\beta$ = 2350 $\text{MPa} \cdot \text{s}^\beta$. In (a), it is evident that the model does not provide a good fit for long-term creep behavior. However, (b) shows that the model fits well within the specified interval.}
\label{Fig06}
\end{figure}

The estimated values of $\beta$ and $C_\beta$, along with the corresponding deformation curves for polypropylene under an applied load of $\sigma_0 = 1.4 \, \text{MPa}$ at $20^\circ \text{C}$, are compared to experimental data across three distinct time intervals in Figs.~\ref{Fig04}, \ref{Fig05}, and \ref{Fig06}. These comparisons highlight the model's precision within limited time frames, reflecting its phenomenological nature.

Literature reports that Eq.(\ref{eq_def_creep}) has been employed effectively to model the creep behavior of elastomers, based on the assumption that both $C_\beta$ and $\beta$ remain constant throughout the experiment \cite{di2011visco}. While this approach may produce accurate results when the experiment's duration is very limited, the above results show this is not the case globally. This limitation is significant because the power-law function utilized in the fractional-order model is not suited for capturing long-term creep behavior, which may evolve differently as time progresses. As time extends beyond the experimental window, the assumptions behind constant parameters may no longer hold, highlighting the need for alternative approaches for long-duration modeling. 

\subsection{The variable-order springpot}

\begin{figure}[h!]
\centering
\includegraphics[width=100mm]{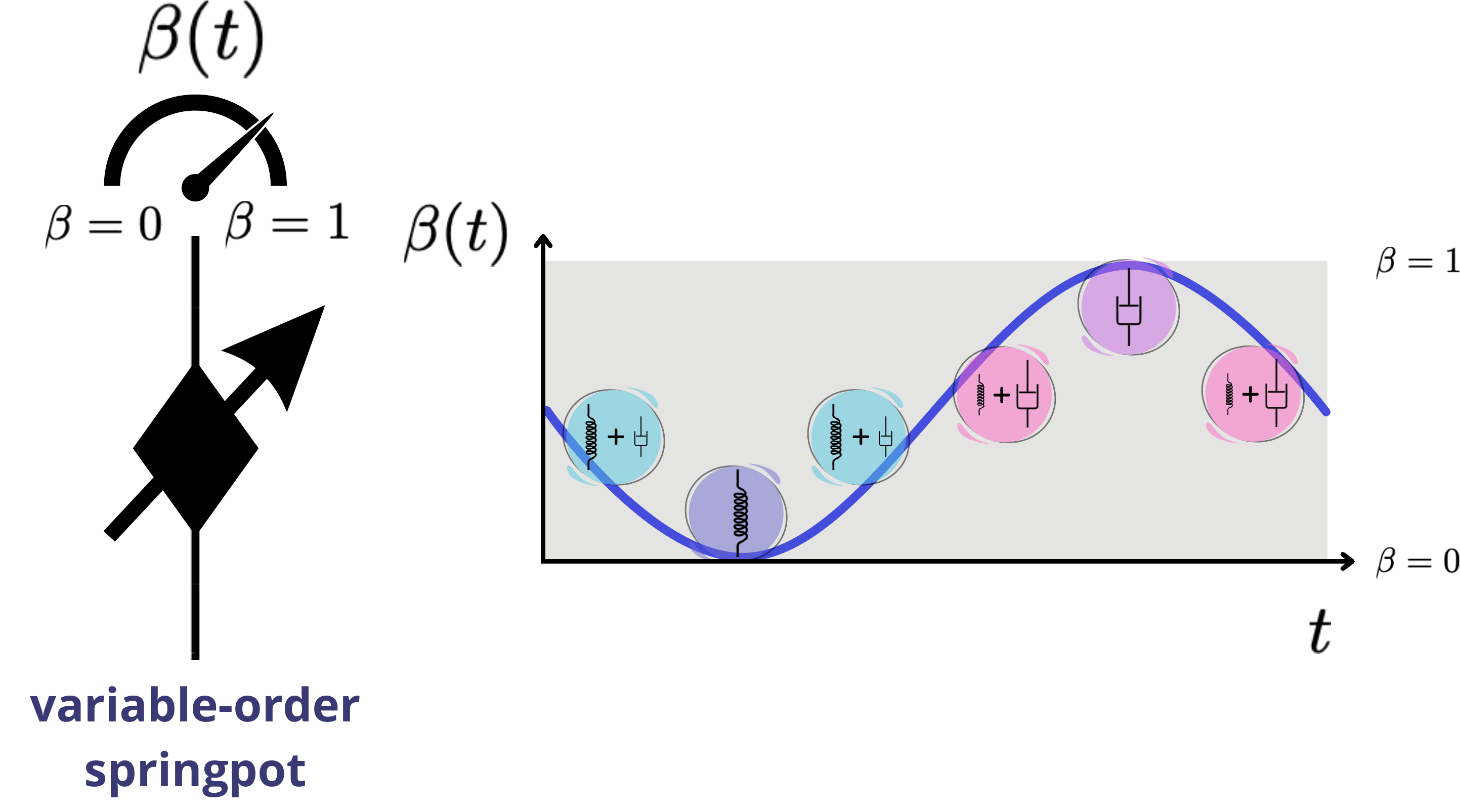}
\caption{Schematic representation of the variable-order springpot model, illustrating its ability to transition between elastic and viscous behaviors over time. The variable-order nature is depicted by $\beta(t)$, which dynamically changes as the material experiences different stages of creep, providing a more accurate representation of polymers's time-dependent mechanical response.}
\label{Fig07}
\end{figure}

The results of the previous section indicate that the rheological model must account for the variation of both parameters, $\beta$ and $C_\beta = T^{\beta} \, E$, which requires specific assumptions. The first assumption is that the parameter $E$, and consequently $T = \eta/E$, depend on the applied stress $\sigma(t) = \sigma_0$, i.e., $E = E(\sigma_0)$ and $T = T(\sigma_0)$. Also, the order parameter must become time-dependent, i.e., $\beta = \beta(t)$. These assumptions together give rise to the \emph{variable-order springpot} shown in Fig~\ref{Fig07}, and the constitutive equation from (\ref{eq_pot_integral_2}) can be written in a variable-order form
\begin{equation}
\varepsilon(t) = \frac{1}{E\left(\sigma_0\right) \, \Gamma[\beta(t)]} \int_{0}^{t}\frac{1}{T\left(\sigma_0\right)} \, \left[ \frac{t}{T\left(\sigma_0\right)} - \frac{\tau}{T\left(\sigma_0\right)} \right]^{\beta(t)-1} 
 \, \sigma_0 \, d\tau \, ,
\label{eq_pot_creep_VO}
\end{equation}
or equivalently
\begin{equation}
\varepsilon(t) = \frac{1}{E\left(\sigma_0\right)} \, \FracInt[g]{}{}{\beta(t)}{\sigma_0} \, .
\label{eq_pot_creep_VO_01}
\end{equation}

Since the parameters $\sigma_0$, $\beta(t)$ and $T\left(\sigma_0\right)$ in Eq.(\ref{eq_pot_creep_VO}) are not a function of $\tau$, and considering the identity $\Gamma[\beta(t)] \, \beta(t) = \Gamma[\beta(t)+1]$, we have that
\begin{equation}
\varepsilon(t) = \frac{\sigma_0}{E\left(\sigma_0\right) \, \Gamma[\beta(t)]} \, \frac{1}{T\left(\sigma_0\right)^{\beta(t)}} \, \int_{0}^{t} \left( t - \tau \right)^{\beta(t)-1} 
 \,  d\tau \ 
\label{eq_pot_creep_VO_02}
\end{equation}
is equivalent to the creep response for a \emph{variable-order springpot} given by
\begin{equation}
\varepsilon(t) = \frac{\sigma_0}{E\left(\sigma_0\right) \, \Gamma[\beta(t) + 1]} \left[ \frac{t}{T\left(\sigma_0\right)} \right] ^{\beta(t)} \, , \quad 0 \leq \beta(t) \leq 1 \, .
\label{eq_pot_creep_VO_03}
\end{equation}

The value of $C_\beta$ depends on the temporal function $\beta(t)$, which, for instance, can be modeled by a function exhibiting a transition from an initial value $\beta_0$  to a final asymptotic value $\beta_\infty$ \cite{garrappa2021variable}, where $0 \leq \beta_0 < \beta_{\infty} \leq 1$. Here the transition function chosen for $\beta(t)$ while dealing with creep is
\begin{equation}
\beta(t) = \frac{\beta_\infty \, \left( \frac{t}{\gamma} \right)^{\delta}+ \beta_0}{ \left( \frac{t}{\gamma} \right)^{\delta} + 1} \, ,
\label{eq_beta_variation}
\end{equation}
where $\gamma > 0$ and $\delta > 0$ are dubbed the \emph{characteristic time} and the \emph{shaping exponent}, respectively. These parameters have distinct physical interpretations:
\begin{itemize}
    \item \textit{Initial fractional order $\beta_0$}: Represents the fractional order at the start of the creep process, corresponding to the material's behavior in the glassy phase. In this phase, $\beta_0$ quantifies the level of initial dissipation.
    \item \textit{Asymptotic fractional order $\beta_{\infty}$}: Denotes the fractional order as the material approaches the rubbery phase. It captures the long-term behavior where the material retains a more pronounced viscoelastic response.
    \item \textit{Characteristic time $\gamma$}: Governs the speed of progression from $\beta_0$ to $\beta_\infty$. Physically, $\gamma$ reflects the rate of molecular chain reorganization, with smaller values indicating faster transitions and larger values corresponding to slower, more gradual changes.
    \item \textit{Shaping exponent $\delta$}: Controls the asymmetry in the transition between $\beta_0$ and $\beta_{\infty}$.
\end{itemize}

\pagebreak
While this study focused on creep behavior, the theoretical framework underlying the variable-order springpot model suggests that it can be extended to capture other types of loading conditions (e.g. relaxation), since other transition functions can be used for $\beta(t)$. Possibly, it will be necessary to use different physical parameters in $\beta(t)$ than those used in creep.

The transition function is selected purely on phenomenological grounds, as it provides a smooth transition governed by a single parameter $\gamma$. However, other functions offering a gradual transition between levels, such as the logistic sigmoid, could be applied without fundamentally altering the model. From a mathematical point of view, any continuous function of time $t$ is an acceptable choice, offering great flexibility to model material with another type of behavior.

Note the variable-order springpot model involves six parameters $(E,\eta,\beta_0, \beta_{\infty}, \gamma, \delta)$, that need to be determined using experimental data. The temporal dependence of $\beta$, encapsulated in Eq.(\ref{eq_beta_variation}), introduces an extra level of complexity into the fractional model, so that now it is not so simple as in Eq.(\ref{eq_def_creep_log}) to estimate the rheological parameters, due to the non-convex nature of the model calibration problem.

To address this, it has been employed an optimization approach using the MATLAB package \textbf{CEopt} (\url{https://ceopt.org}), which leverages the Cross-Entropy (CE) method \cite{CunhaJr2024CEopt}. The CE method is particularly suitable for this application as it transforms the non-convex optimization problem into a rare event estimation problem, solvable with Monte Carlo simulations \cite{kroese2013handbook, cunha2014uncertainty}. Unlike gradient descent, which may get trapped in local minima, or genetic algorithms, which can be computationally demanding, CE is robust in handling non-convex landscapes, particularly in mechanics applications \cite{cunha2021enhancing,Issa2023cobem,Dantas2019cobem,Dantas2019icedyn}

The CE method is chosen here for its robustness in exploring parameter spaces and avoiding premature convergence on local minima. It iteratively refines its search, focusing on high-performance areas within the parameter space, which is essential when working with complex, non-linear models like the variable-order springpot. This process allows CE to achieve a stable, physically consistent calibration that is less dependent on initial parameter guesses.

\begin{figure}[h!]
\centering
\includegraphics[width=100mm]{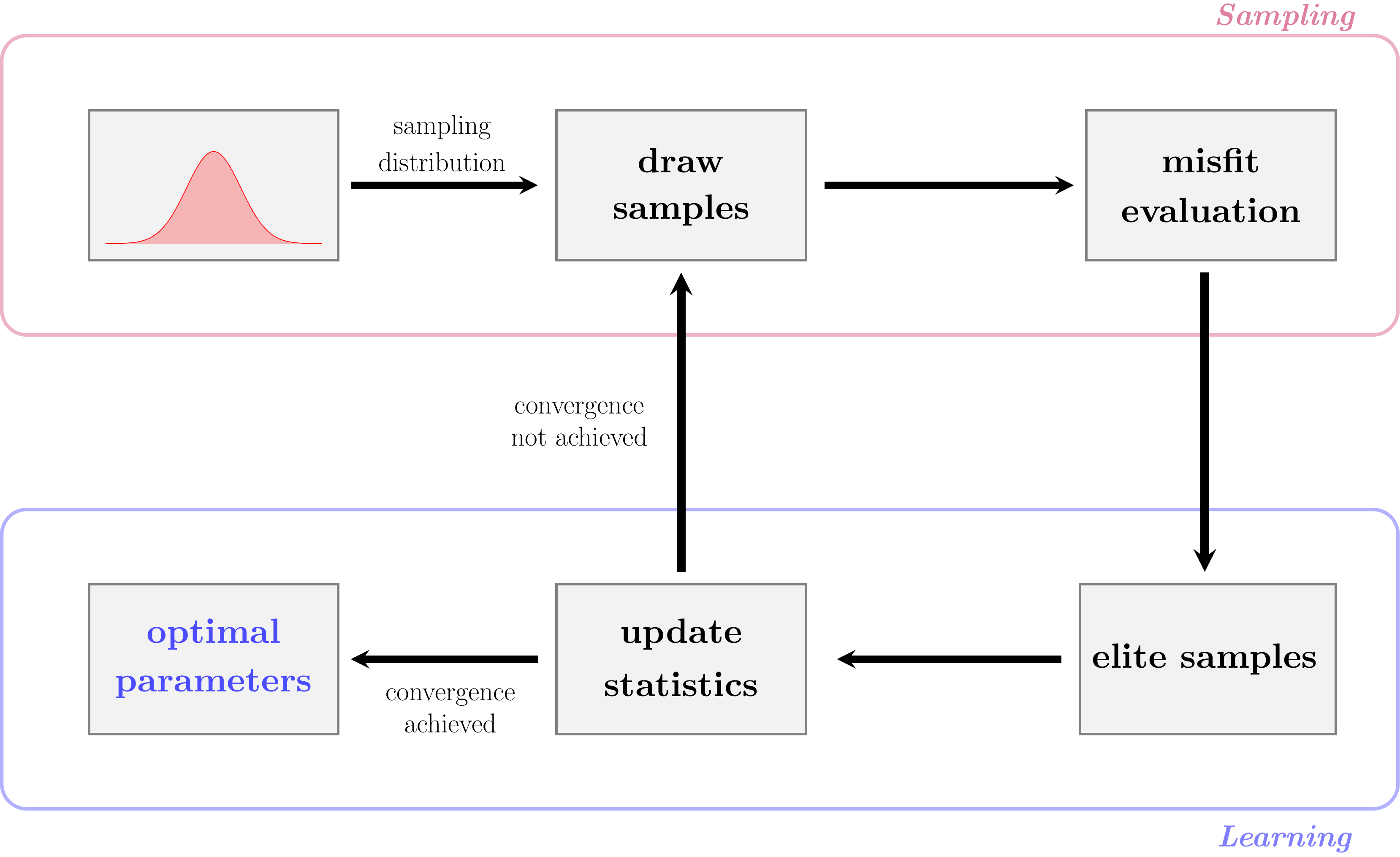}
\caption{ Flowchart of the cross-entropy (CE) optimization process used for model calibration. The process begins by sampling parameter values from an initial distribution, followed by drawing candidate samples. Each sample's misfit is evaluated to determine its fitness relative to experimental data. A subset of elite samples is selected to update the parameter distribution for the next iteration. The process repeats until convergence criteria are met, producing optimal parameter values that yield physically consistent model fittings. This robust method ensures convergence towards realistic parameter values, provided that the initial search region is well-defined.}
\label{Fig08}
\end{figure}

\textbf{Challenges in Parameter Estimation:} A key challenge in parameter estimation for this model is ensuring that the selected parameters remain within physically meaningful ranges, as deviations could lead to unrealistic model behavior. Additionally, the non-convexity of the problem adds a layer of difficulty in achieving convergence. By defining a carefully bounded initial search space based on physical admissibility, the CE method can produce reliable parameter estimates that adhere to material properties. The CE algorithm’s steps are summarized in Fig.~\ref{Fig08} and outlined below for clarity:

\begin{enumerate}
    \item \emph{Sampling from Initial Distributions:} Define a broad initial distribution for each parameter within a physically admissible range, reflecting prior knowledge of the parameters and ensuring consistency with the material's behavior.
    \item \emph{Generating Samples:} Draw a set of candidate solutions (samples) from the initial parameter distributions, representing potential values for each model parameter.
    \item \emph{Evaluating the Misfit Function:} For each sample, calculate the misfit function, quantifying the difference between model predictions and experimental data. Lower misfit values indicate better agreement with observations.
    \item \emph{Selecting Elite Samples:} Identify a subset of top-performing samples, called \emph{elite samples}, based on their misfit values. These elite samples guide the optimization toward promising parameter regions.
    \item \emph{Updating Distribution Hyperparameters:} Update the distribution parameters (mean and standard deviation) based on elite samples, focusing subsequent searches on regions of lower misfit values.
    \item \emph{Checking for Convergence:} Iterate the sampling and updating steps until convergence, which occurs when successive iterations yield minimal improvement in misfit or when the standard deviation plateaus.
    \item \emph{Outputting Optimal Parameters:} Upon convergence, the algorithm outputs the optimal values of $(E,\eta,\beta_0, \beta_{\infty}, \gamma)$, providing a physically consistent calibration that accurately reflects the model's behavior.
\end{enumerate}

The proposed misfit function for the model calibration problem is defined as
\begin{equation}
\mathcal{J}(E,\eta,\beta_0, \beta_{\infty}, \gamma, \delta) = \sqrt{\frac{\displaystyle \sum _{i=1}^{N} \left(\varepsilon_{exp}^i - \varepsilon^i(E,\eta,\beta_0, \beta_{\infty}, \gamma, \delta) \right)^2}{\displaystyle  \sum _{i=1}^{N} \left(\varepsilon_{exp}^i \right)^2}} \times 100 \, ,
\label{eq_MinFunc}
\end{equation}
where $N$ is the number of points sampled from the curves, $\varepsilon_{exp}^i$ are the values sampled from experimental curve at instants $t_i$, $\varepsilon^i$ are the predicted values from Eq.(\ref{eq_pot_creep_VO_03}). The misfit value $\mathcal{J}$ is a measure of the percentage deviation between theoretical and experimental curves.

\paragraph{Why a \emph{single} variable–order springpot?} Classical generalisations of the Kelvin--Voigt or Maxwell models replace \emph{both} the spring and the dashpot with fixed–order springpots \cite{jozwiak2015fractional,bonfanti2020fractional}. Although two–springpot assemblies can capture the coexistence of two power–law regimes, they still require six \emph{independent and time–\emph{invariant}} parameters. Consequently, they portray the glassy--to--rubbery transition as an \emph{abrupt} switch between two pre–assigned exponents, while experimental evidence shows that the material stiffness evolves \emph{continuously} during creep. The present approach follows a different philosophy: we keep a single springpot but promote its order to a \emph{time--dependent} quantity $\beta(t)$, augmented here by the shape exponent $\delta$ in Eq.~\eqref{eq_beta_variation}. This choice adds just one extra scalar to the parameter set yet endows the model with an entire continuum of intermediate viscoelastic states. In other words, the variable–order framework attains the descriptive power usually associated with multiple elements \emph{without} the proliferation of static parameters, preserves physical interpretability ($E$ and $\eta$ retain their classical meaning) and yields a calibration problem that remains well posed throughout the explored stress range. For these reasons—parsimony, smooth representation of phase evolution, and parameter identifiability—we deliberately limited the analysis to a single variable–order springpot, deferring multi–element extensions to future work focused on, for example, non–isothermal loading or cyclic fatigue.

\section{Results and Discussion}

This section evaluates the effectiveness of the proposed variable-order fractional model in replicating the creep behavior of PP and PVC, based on experimental data from the literature \cite{castro2016fatigue,crawford2020plastics}. The data detail the creep response of PP under loads ranging from 1.4 MPa to 14 MPa at a constant temperature of 20°C and for PVC at 20°C under loads ranging from 5 MPa to 40 MPa. The estimated parameter values of Eqs.(\ref{eq_pot_creep_VO_03})-(\ref{eq_beta_variation}) for PP and PVC are provided in Tab.~\ref{Tab_Parameters}, where the misfit values indicate strong alignment between the model predictions and experimental data.

\begin{table}[ht]
\caption{Estimated parameter values for modeling the creep behavior of PP and PVC at 20°C under various applied stresses ($\sigma_0$). The parameters $E$ (elastic modulus), $\eta$ (viscosity), $\beta_0$ (initial fractional order), $\beta_\infty$ (asymptotic fractional order), $\gamma$ (transition rate) and $\delta$ (shaping exponent) were calibrated for each stress level. Misfit values indicate the accuracy of model fit to the experimental data, demonstrating good adherence across stress conditions.}
\label{Tab_Parameters}
\centering
\begin{tabular}{c@{}ccccccccc}
\toprule
Material & ~~~ $\sigma_0$ \, (MPa) & $E$ \, (GPa) & $\eta$ \, (GPa . s)  & $\beta_0$ & $\beta_\infty$ & $\gamma$ (s) & $\delta$ & Misfit (\%)\\
\midrule
        & 2.8  & 1.25 & 222.36 &  0.0363 & 0.1015 & 4403 & 0.70 & 0.81\\
        & 4.2  & 1.22 & 281.20 &  0.0397 & 0.1005 & 3758 & 0.70 & 0.83\\
        & 5.6  & 1.18 & 200.01 &  0.0475 & 0.1030 & 3113 & 0.70 & 0.77\\
  PP    & 9.8  & 1.02 & 200.01 &  0.0687 & 0.1147 & 1200 & 0.70 & 0.98\\
        & 11.2 & 0.96 & 200.06 &  0.0720 & 0.1216 & 1200 & 0.70 & 0.27\\
        & 12.6 & 0.87 & 265.97 &  0.0866 & 0.1258 & 1200 & 0.70 & 0.27\\
\midrule
        & 10 & 2.62 & 278.69 &  0.0148 & 0.0689 & 36578 & 0.58 & 0.14\\
        & 20 & 2.21 & 302.52 &  0.0260 & 0.0750 &  9431 & 0.42 & 0.46\\
  PVC   & 25 & 1.88 & 304.88 &  0.0291 & 0.1091 &  6000 & 0.26 & 0.95\\
        & 35 & 1.08 & 213.86 &  0.0611 & 0.1774 &  2400 & 0.22 & 0.49\\
\bottomrule
\end{tabular}
\vspace*{-4pt}
\end{table}

In typical loading scenarios, each of the parameters in Eqs.(\ref{eq_pot_creep_VO_03}) and (\ref{eq_beta_variation}) depends on the applied load $\sigma_0$, a functional dependence that may be well captured through
\begin{equation}
\left\{ E, \eta, \beta_0, \beta_{\infty}, \gamma, \delta \right \} = \frac{ \left( m_1 \, \sigma_0 + m_2 \right) \left( \cfrac{\sigma_0}{\sigma_r}  \right ) ^{n} +  m_3 \, \sigma_0 + m_4}{\left( \cfrac{\sigma_0}{\sigma_r}  \right )^{n} + 1} \, ,
\label{eq_parameters}
\end{equation}
a phenomenological equation that adapts to very diverse changes in mechanical behavior. Here $\sigma_r$ is a reference stress level, $n$ is a kind of shape parameter, while $m_1$, $m_2$, $m_3$, and $m_4$, assume different interpretations depending on the physical quantity being fitted. Further details about this equation can be seen in the Supplementary Material.

The viscosity parameter $\eta$ reflects the chain‐to‐chain (microscopic) friction in the polymer, which (for the temperature under analysis) is expected to be essentially stress‐independent as long as the applied load stays well below the yield strength. In the present tests (1.4 -- 14 MPa for PP and 5 -- 40 MPa for PVC) no experimental or theoretical mechanism predicts a genuine change of $\eta$. The mild “trend’’ observed in the fit of Table~\ref{Tab_Parameters} therefore stems from over-fitting: the optimizer varied $\eta$ to compensate for the limited shape freedom of the former $\beta(t)$ law. We now fix $\eta$ at the smallest values that still reproduce every creep curve, namely $\eta=300 \text{GPa.s}$ for PP and $\eta=300 \times 10^3 \text{GPa.s}$ for PVC, thereby eliminating the ill-condition and stabilizing the remaining calibration parameters.

The estimated parameter values for PP and PVC phenomenological equations are provided in Tab.~\ref{Tab_Equations_PP}. To obtain this fit, the variable-order fractional model was first calibrated according to the procedure described in section~\ref{sec3}, for several values of stress level $\sigma_0$. The values of $(E, \eta, \beta_0, \beta_{\infty}, \gamma, \delta)$ resulting from this sequence of curve fittings were used in a secondary calibration process with the aid of Eq.(\ref{eq_parameters}) to produce the results shown in the Tab.~\ref{Tab_Equations_PP}. Notably, the results indicate these parameters present functional dependence on $\sigma_0$, in agreement with the hypothesis used in the theoretical development of the variable-order springpot.

\begin{table}[ht]
\caption{Estimated parameter values for modeling the creep behavior of PP and PVC at 20°C under various applied stresses ($\sigma_0$). The parameters $E$ (elastic modulus), $\eta$ (viscosity), $\beta_0$ (initial fractional order), $\beta_\infty$ (asymptotic fractional order), and $\gamma$ (characteristic time) and $\delta$ (shaping exponent) were calibrated for each stress level $\sigma_0$.}
\label{Tab_Equations_PP}
\centering
\begin{tabular}{ccccccccl}
\toprule
    Material & Parameter & $\sigma_r$ \, (MPa) & $n$ & $m_1$ & $m_2$ & $m_3$ & $m_4$ & Unit\\
\midrule
    & $E$            & 11.72 & 2.63  & -53.20 & 1415.24 &  -21.28 & 1306.29         & MPa\\
    & $\eta$         & ----- & ----- &      0 &       0 &       0 & 3 $\times 10^2$ & GPa . s\\
    & $\beta_0$      & ----- & ----- &      0 &       0 &   0.005 & 0.0203          & ----\\
PP  & $\beta_\infty$ & 10.43 &  5.49 &      0 & 0.1347  &       0 & 0.1012          & ----\\
    & $\gamma$       &  7.60 & 15.56 &      0 & 1200.54 & -454.03 & 5669.17         & s\\
    & $\delta$       & ----- & ----- &      0 &       0 &       0 & 0.70            & -----\\
\midrule
     & $E$             & 21.24 & 3.27  & -76.20 & 3587.29 &   -27.38 &  2879.71        & MPa\\
     & $\eta$          & ----- & ----- &      0 &       0 &        0 & 3 $\times 10^5$ & GPa . s\\
     & $\beta_0$       & 21.81 & 9.33  & 0.0036 & -0.0659 &   0.0020 & -0.005          & ----\\
 PVC &  $\beta_\infty$ & 26.54 & 10.34 &      0 &  0.1835 &        0 &  0.0690         & ----\\
     & $\gamma$        & 14.94 & 3.70  &      0 & 2319.31 & -1260.46 &  56959.33       & s\\
     & $\delta$        & 23.16 & 12.62 &      0 &    0.22 &  -0.0130 &  0.708          & ----\\
\bottomrule
\end{tabular}
\vspace*{-4pt}
\end{table}

\begin{figure}[ht]
\centering
\includegraphics[width=130mm]{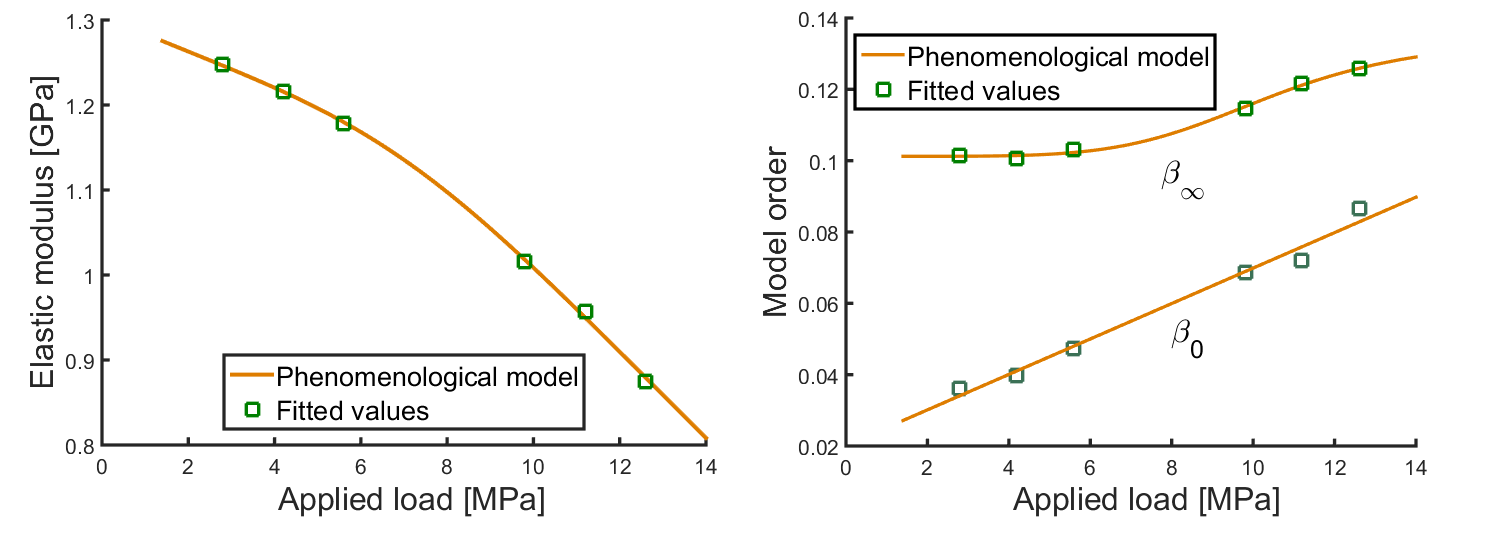}
\caption{PP phenomenological curves for the parameters $E$, $\beta_0$ and $\beta_\infty$ as function of the applied load $\sigma_0$. The curves illustrate the sensitivity of each parameter to stress levels, suggesting that higher loads affect elasticity and fractional order differently.}
\label{Fig09}
\end{figure}

\begin{figure}[ht]
\centering
\includegraphics[width=130mm]{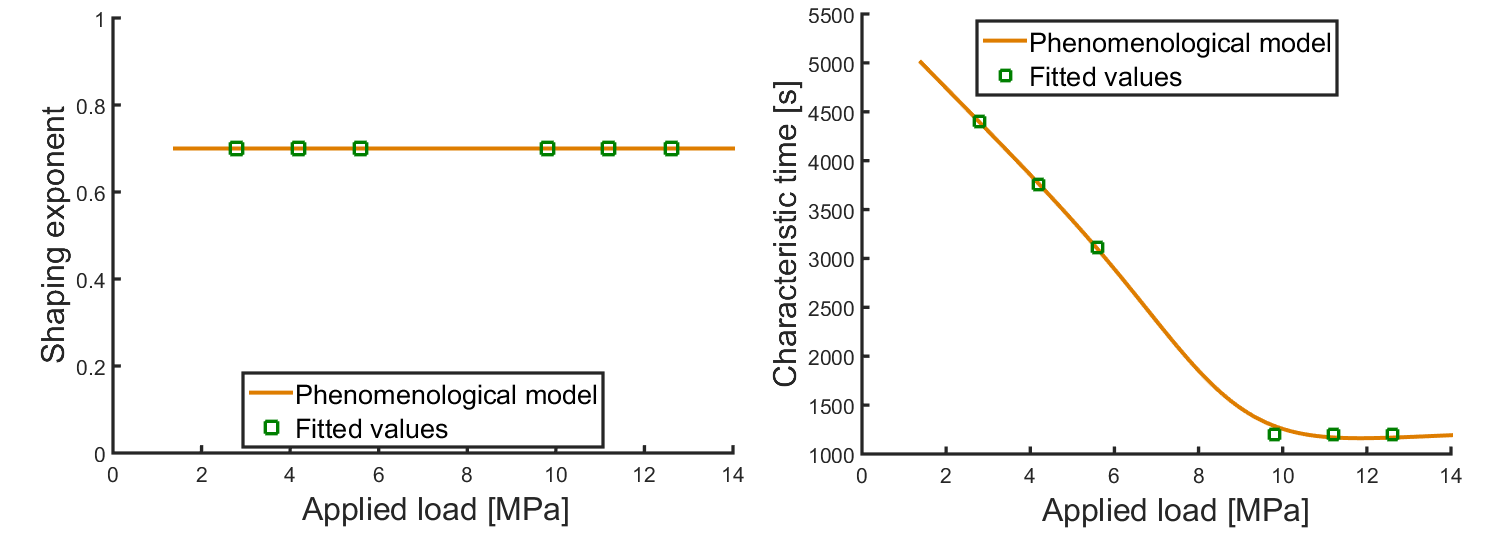}
\caption{PP phenomenological curves for $\delta$ and $\gamma$ as functions of applied load $\sigma_0$. The parameter $\delta$ shows constant behavior. Conversely, $\gamma$ decreases linearly until a saturation point after $\sigma_0 = 10$ MPa. This nonlinear behavior illustrates the complex stress-dependent dynamics of polypropylene.}
\label{Fig10}
\end{figure}

The behavior of $E$, $\beta_0$, and $\beta_\infty$ in relation to load $\sigma_0$ for PP are illustrated in Fig.~\ref{Fig09}, while $\delta$ and $\gamma$ are shown in in Fig.~\ref{Fig10}. Since $\eta$ is assumed constant with $\sigma_0$ for both materials the curves are not shown. These results highlight the observed behavior of the parameters in the proposed variable-order rheological model of Eq.(\ref{eq_pot_creep_VO_03}). As shown in Fig.~\ref{Fig09}, the elastic modulus $E$ decreases with increasing applied stress $\sigma_0$, contributing to a reduction in stiffness. The fractional order parameters, $\beta_0$ and $\beta_\infty$, increase with $\sigma_0$, capturing the evolving time-dependent characteristics of the material. Figure~\ref{Fig10} illustrates that the shaping exponent $\delta$ of PP does not change with an increasing stress $\sigma_0$. Conversely, the transition rate parameter $\gamma$ decreases with $\sigma_0$, indicating faster transitions between viscoelastic phases at higher loads until a saturation after $\sigma_0 = 10$ MPa. The isolated behavior of these parameters does not directly describe the global behavior of PP under load, as their interplay is balanced by the proposed variable-order rheological model to accurately capture the material's complex creep behavior.

\begin{figure}[ht]
\centering
\includegraphics[width=130mm]{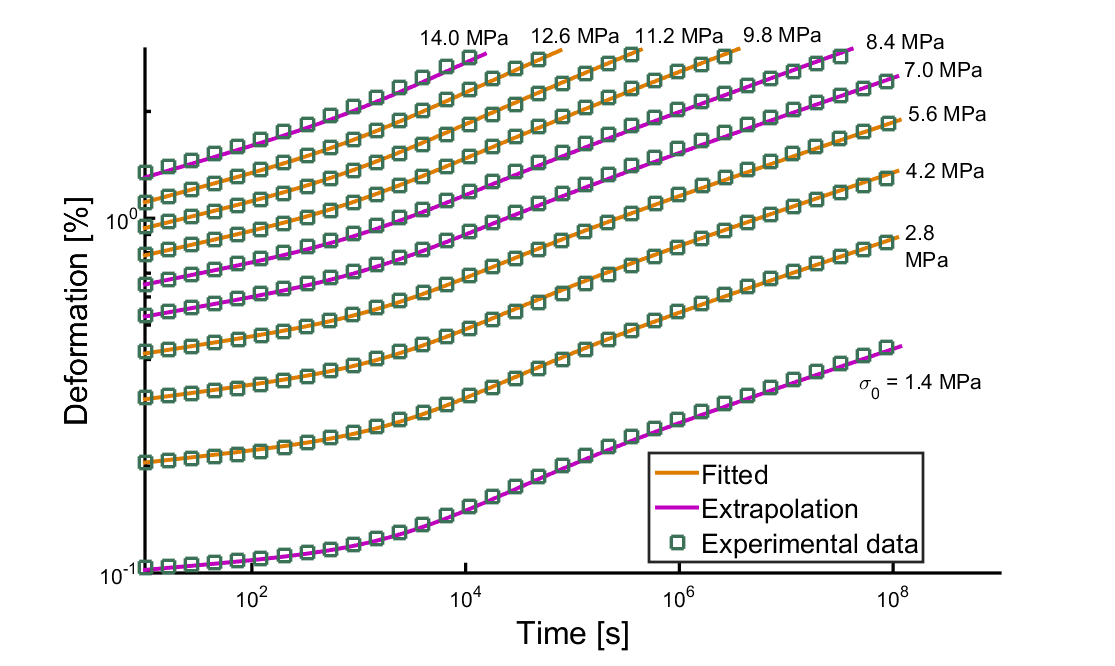}
\caption{Comparison of experimental and modeled creep curves for PP at $20^\circ \text{C}$ under constant stress levels between $1.4$ and $14 \, \text{MPa}$. The variable-order fractional model successfully reproduces the experimental data, accurately capturing the evolution of creep deformation across different stress levels. At lower loads, the model aligns well with experimental observations, reflecting PP's initially slow deformation in the glassy phase. As the load increases, the model also accommodates the accelerated creep behavior associated with the material’s transition to a more viscoelastic state. In the rubbery phase, the model demonstrates its flexibility in describing long-term deformation behavior where steady-state creep is observed. The close fit between the model predictions and experimental data across these phases highlights the variable-order model's ability to capture the complex time-dependent mechanical response of PP under sustained load, making it a valuable tool for predicting long-term material performance in practical applications.}
\label{Fig11}
\end{figure}

Together, Eqs.~(\ref{eq_pot_creep_VO_03}), (\ref{eq_beta_variation}), and (\ref{eq_parameters}) establish a system that reliably models the creep behavior of PP at 20°C for applied loads between 1.4 and 14 MPa. This system combines both variable-order fractional calculus and phenomenological modeling to capture the complex viscoelastic response of polypropylene. The comparison in Fig.~\ref{Fig11} between the model's calculated deformation curves and the experimental data highlights the model's effectiveness, with close alignment over a broad range of applied stresses. Notably, the model accurately describes the deformation trajectory not only for stress levels used during calibration but also for unseen conditions, including stress levels within the training domain (e.g., 7.0 or 8.4 MPa) and at its boundaries (1.4 MPa and 14 MPa). These results highlight the model's predictive capability and robustness in describing the time-dependent mechanical behavior of PP under various loading conditions.

\begin{figure}[h!]
\centering
\includegraphics[width=130mm]{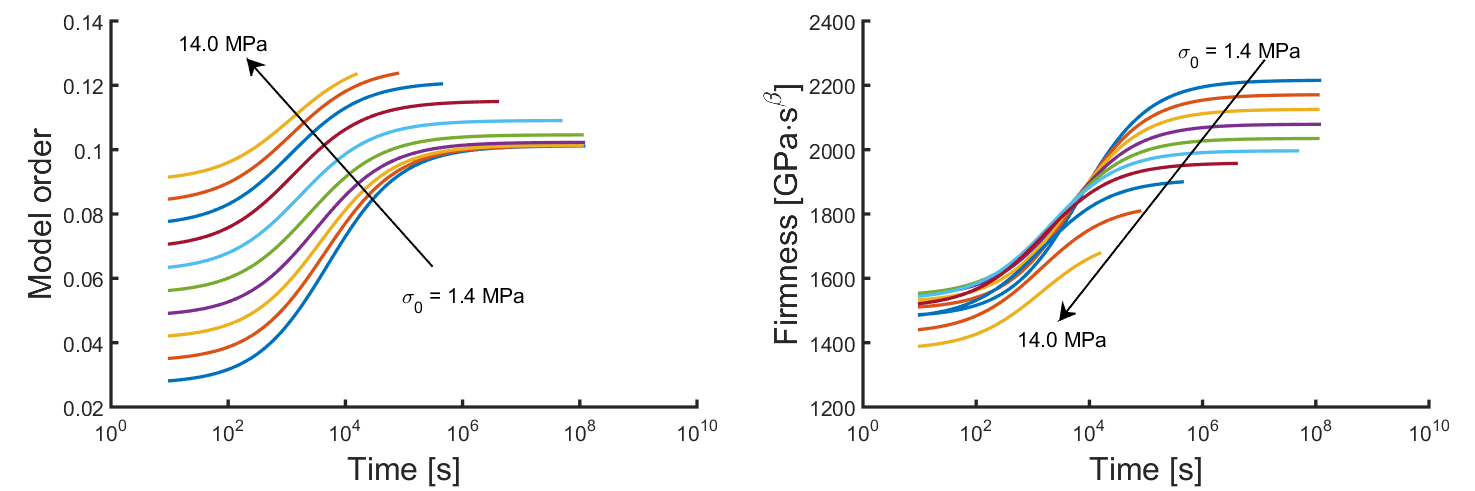}
\caption{Evolution of the order parameter $\beta$ and firmness $C_\beta$ for PP at $20^\circ \text{C}$ under constant stress levels between $1.4$ and $14 \, \text{MPa}$. Initially, $\beta$ remains stable in the glassy phase, where PP exhibits predominantly elastic behavior. As time progresses, $\beta$ increases, signaling a gradual transition to a more dissipative, viscoelastic response. The firmness parameter $C_\beta$, which reflects the material’s resistance to deformation, shows a complementary trend: initially, it aligns closely with the elastic modulus but increases during the transition phase, indicating enhanced molecular mobility and progressive loss of stiffness. Eventually, both parameters stabilize, with $\beta$ reaching an asymptotic value characteristic of the rubbery phase, while $C_\beta$ converges to a plateau, representing a steady-state balance between elasticity and viscosity. These trends demonstrate the variable-order fractional model’s ability to capture the complex interplay between stress, molecular rearrangement, and viscoelastic phase transitions in PP.}
\label{Fig12}
\end{figure}

The variable-order nature of $\beta(t)$, governed by Eq.~(\ref{eq_beta_variation}), introduces flexibility to the model by allowing the material's behavior to shift progressively over time. This shift aligns with the physical understanding that polymers like PP undergo microstructural changes over time, such as molecular chain realignment, that influence mechanical properties.

Initially, as shown in Fig.~\ref{Fig12} left, $\beta(t)$ remains stable in the glassy phase, where PP exhibits predominantly elastic behavior, resisting deformation with minimal energy dissipation. As time progresses and the polymer enters the transition phase, $\beta(t)$ begins to increase, marking the onset of a more dissipative (viscous) response, where molecular mobility facilitates deformation under sustained load. In the rubbery phase, $\beta(t)$ stabilizes at a higher asymptotic value ($\beta_\infty$), reflecting a gradual approach to a stable viscoelastic state.

Despite this gradual approach to a more viscous-like state, the observed $\beta(t)$ values remain relatively small, ranging from slightly less than 0.04 to around 0.12. This narrow range of small values means that the firmness parameter $C_\beta$, governed by $C_\beta = \eta^{\beta} \, E^{1-\beta}$, has a much stronger dependence on $E$ than on $\eta$. As a result, the interplay between elasticity and viscosity in the observed time-dependent mechanical behavior of PP is dominated by the elastic component, even in the rubbery phase. This dynamic variation of $\beta(t)$ underscores the model's capability to capture the complex, evolving viscoelastic behavior of PP as it transitions across phases while maintaining its predominantly elastic character.

Similarly, Fig.~\ref{Fig12} right shows how the firmness parameter $C_\beta$ evolves under varying stress conditions and over time. Initially, in the glassy phase, $C_\beta$ closely approximates the modulus of elasticity, indicating that PP primarily resists deformation through elastic forces. As time progresses and the material transitions into the viscoelastic phase, $C_\beta$ increases, reflecting a gradual reduction in stiffness due to molecular chain mobility. However, while some viscous behavior emerges, the material's response remains predominantly elastic. Furthermore, $C_\beta$ decreases with increasing applied load $\sigma_0$, indicating greater compliance at higher loads. This trend is consistent with the observed reduction in $E$ with increasing $\sigma_0$ (Fig.~\ref{Fig09}). Once PP reaches the rubbery phase, $C_\beta$ stabilizes, signifying a steady state where elastic resistance continues to dominate, even as some viscous flow occurs. These trends emphasize that elasticity remains a central component of PP’s time-dependent mechanical response, even under varying conditions of stress and over long time scales.

\begin{figure}[h]
\centering
\includegraphics[width=130mm]{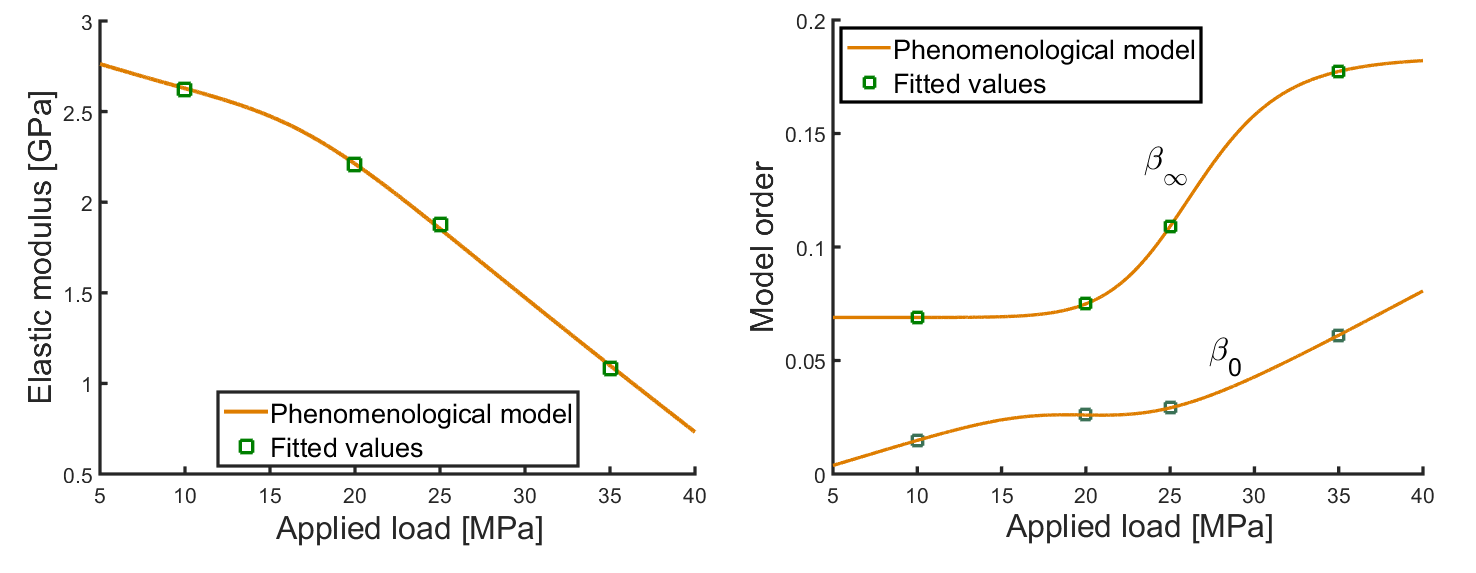}
\caption{PVC phenomenological curves for the parameters $E$, $\beta_0$ and $\beta_\infty$ as function of the applied load $\sigma_0$. The sensitivity of the parameters to stress levels suggests that higher loads affect elasticity and fractional order differently.}
\label{Fig13}
\end{figure}

\begin{figure}[h]
\centering
\includegraphics[width=130mm]{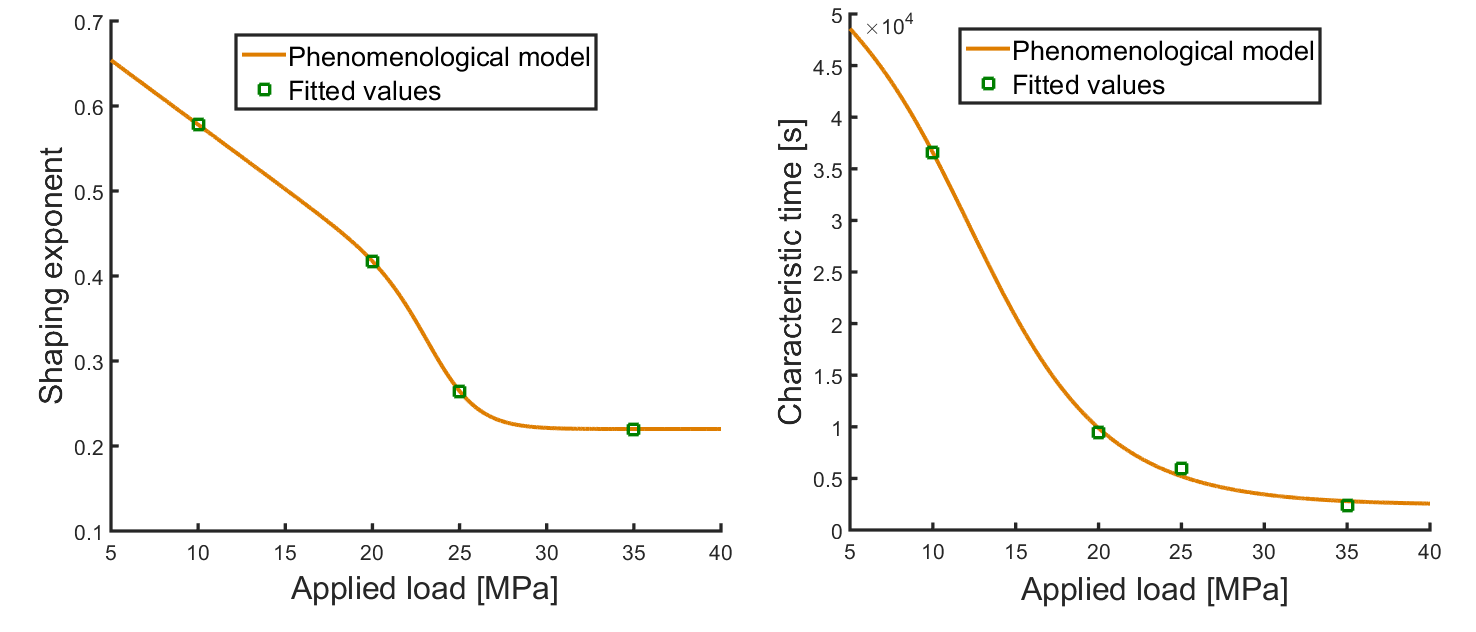}
\caption{PVC phenomenological curves for $\delta$ and $\gamma$ as functions of applied load $\sigma_0$. Both parameters show an initial decreasing tendency followed by a saturation, a complex stress-dependent nonlinear behavior.}
\label{Fig14}
\end{figure}

\begin{figure}[ht]
\centering
\includegraphics[width=130mm]{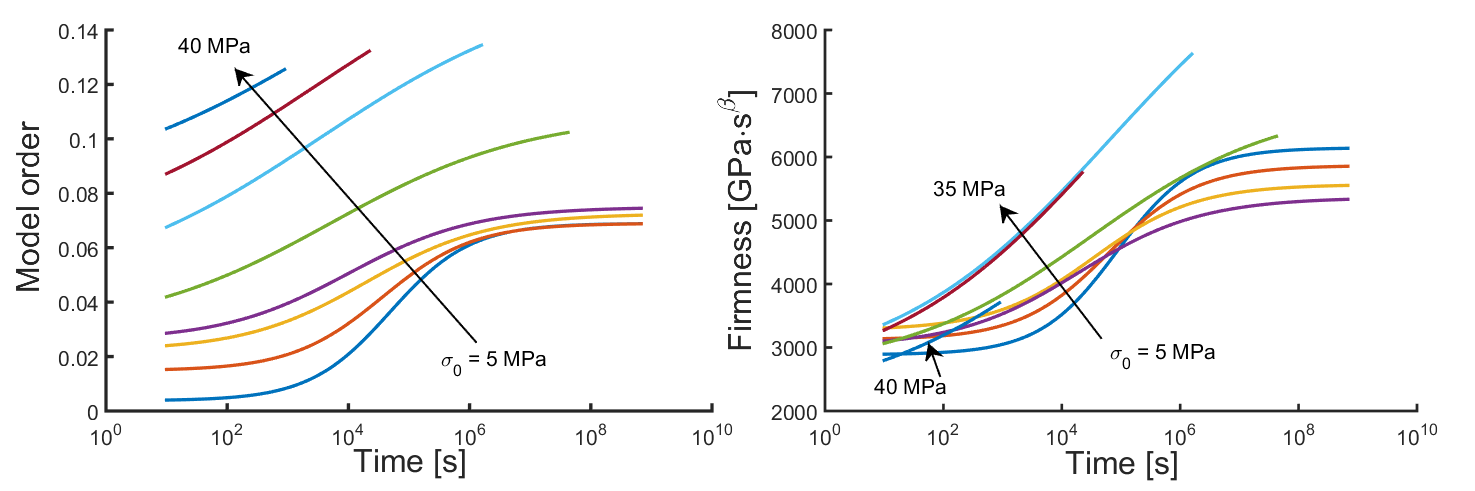}
\caption{Evolution of the order parameter $\beta$ and firmness $C_\beta$ for PVC at $20^\circ \text{C}$ under constant stress levels between $5$ and $40 \, \text{MPa}$. The model order $\beta$ exhibits a monotonic increase over time for all stress levels, reflecting a progressive shift from an elastic-dominated response to a more dissipative, viscoelastic behavior. Higher applied stresses lead to higher initial values and asymptotic limits of $\beta$, indicating an accelerated transition toward a more viscous state. The firmness $C_\beta$ displays a complex, non-monotonic evolution. Initially, $C_\beta$ remains nearly constant. This is followed by a stress-dependent bifurcation in behavior: for lower stresses, $C_\beta$ increases gradually over time, indicating a steady but moderate stiffening effect. In contrast, for higher stresses, the stiffness increases significantly in later stages.}
\label{Fig15}
\end{figure}

\begin{figure}[h]
\centering
\includegraphics[width=130mm]{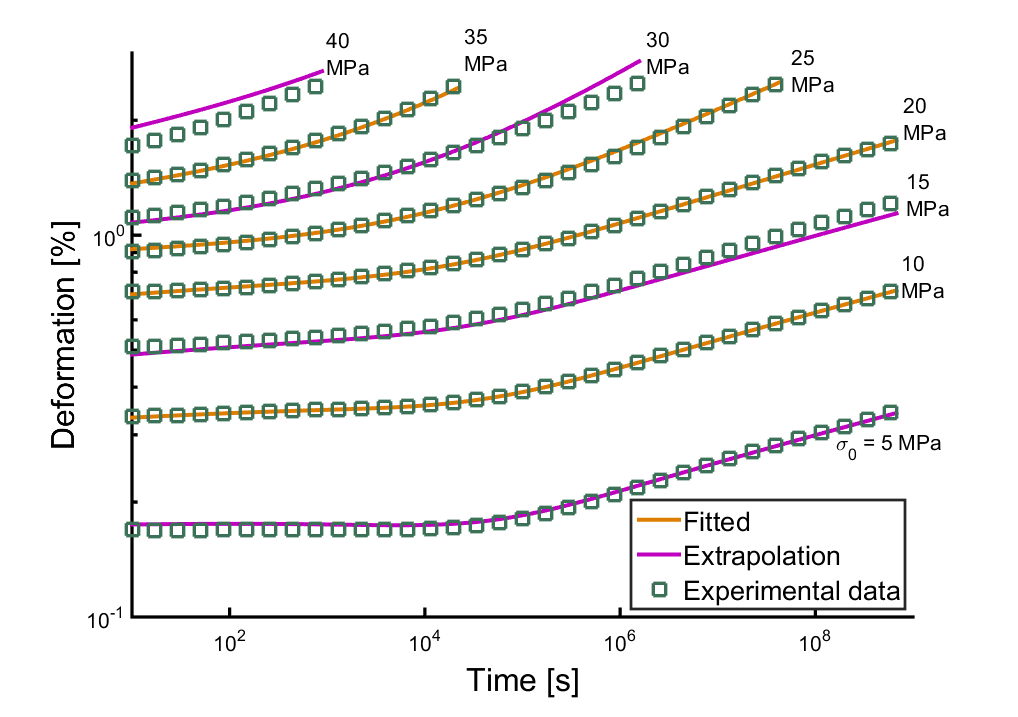}
\caption{Comparison of experimental and modeled creep curves for PVC at $20^\circ \text{C}$ under constant stress levels between $5$ and $40 \, \text{MPa}$. The variable-order fractional model successfully reproduces the experimental data, accurately capturing the evolution of creep deformation across different stress levels.}
\label{Fig16}
\end{figure}

Phenomenological curves similar to those presented in Figs.~\ref{Fig09}, \ref{Fig10}, and \ref{Fig12} were also obtained for PVC, as shown in Figs. \ref{Fig13}, \ref{Fig14}, and \ref{Fig15}. Using the same methodology employed to generate deformation curves for PP, we computed the corresponding deformation results for PVC, which are presented in Fig.~\ref{Fig16}. Here, for PVC, it is important to highlight that $\delta$ is not a constant function of $\sigma_0$, and $\beta_{\infty}$ has a stronger tendency of growth in comparison with PP results shown above.

The deformation curves for PVC in Fig.~\ref{Fig16} show good agreement between model predictions and experimental data, validating the parameter calibration. While minor deviations appear at long timescales and high stresses, likely due to unmodeled effects, the model remains robust and effective for predicting PVC creep behavior.

By successfully applying the model to a second material, PVC, we demonstrate its generality in capturing the viscoelastic behavior of different polymeric materials. These findings reinforce the versatility of the proposed variable-order springpot approach, suggesting that it can be extended to other viscoelastic materials beyond polypropylene, while maintaining its ability to accurately predict time-dependent mechanical responses under varying stress conditions.

While the present study focuses on creep loading, the governing equations are readily extendable to other histories (relaxation, cyclic, thermo-viscoelasticity) by exploiting the variable-order kernel. Validation under such conditions is left for future work.


\section{Conclusion}

This study presented a novel variable-order fractional model to predict the creep behavior of polymers at room temperature under several applied loads. Through the application of variable-order fractional calculus, the model effectively captures the evolving viscoelastic characteristics of PP and PVC across different stress levels, offering a significant advantage over traditional models that rely on constant parameters. By allowing the order of the fractional derivative, $\beta$, to vary over time, this approach reflects the material's gradual transition from elastic-dominant to viscous-dominant behavior during the creep process. The model is grounded in fractional viscoelastic theory rather than empirical regression; however, its predictive capability has so far been demonstrated only for creep. Ongoing efforts are directed toward generalising and validating the formulation for arbitrary loading programmes.

A key feature of this model is the assumption that elasticity $E$ and viscosity $\eta$ remain constant throughout creep for a fixed stress load $\sigma_0$, with the progressive increase in $\beta$ — approaching an asymptotic limit $\beta_\infty$ — governing the material’s response as it transitions through glassy, transition, and rubbery phases. Additionally, the model introduces an initial fractional order $\beta_0$, which is dependent on the applied stress $\sigma_0$ and suggests that some viscous elements become active immediately upon load application. This reflects a nuanced understanding of polypropylene's microstructural response under initial loading conditions and marks a departure from fixed-order rheological models.

Despite the expanded capability of the novel rheological model in comparison with the open literature, limitations were encountered, particularly regarding the model's reliance on phenomenological expressions for parameter calibration. While these expressions provide a good fit, the calibration process could benefit from additional experimental data across more varied temperature and loading conditions to further enhance the model's accuracy and applicability. Additionally, for a fixed load $\sigma_0$, the model assumes constant values for $E$ and $\eta$, which may not fully capture the behavior of materials that undergo significant changes in elasticity and viscosity under varying environmental factors or prolonged stress.

Future extensions of this work could explore alternative models such as those involving multiple springpots (e.g. variable-order Maxwell like), the adaptability of the variable-order springpot model to other ranges of temperature for viscoelastic materials that exhibit similar time-dependent behaviors or even other classes of materials subjected to creep (e.g. composite). By adjusting the parameters to reflect the properties of different materials, the model may be applicable to a wide range of creeping materials, thus expanding its relevance in materials science.

Additionally, an intriguing avenue for further study involves investigating the fundamental limits of the model parameters as imposed by thermodynamical constraints, particularly the second law of thermodynamics. By examining how the model aligns with the principles of energy dissipation and entropy generation, this approach could yield insights into the feasible range and interdependence of parameters like $\eta$, $\gamma$, $\beta_0$, $\beta_{\infty}$, and $\delta$ ensuring that the model’s predictions remain physically consistent across varying conditions. This thermodynamic perspective would not only refine the parameter space but also provide a framework to develop more robust models grounded in physical laws, thereby enhancing their predictive power and applicability to a broader spectrum of materials.

In conclusion, the variable-order fractional model offers a significant advancement over traditional approaches by dynamically linking stress-dependent parameters to time-dependent deformation. This adaptability is particularly beneficial for applications requiring precise long-term predictions of creep deformation in load-bearing polypropylene components, ensuring structural integrity and performance. By highlighting the material’s progression across different viscoelastic phases, the model not only contributes to a deeper understanding of polypropylene’s mechanical behavior but also sets a foundation for future studies to expand the utility of variable-order fractional models in viscoelastic modeling.

\vskip6pt

\enlargethispage{20pt}


\dataccess{The results reported here were obtained with the code \textbf{SpringpotTune: Variable-Order Springpot Model Tuning via Cross-Entropy Method}, which is available at repositories on GitHub (\url{https://americocunhajr.github.io/SpringpotTune}) and Zenodo (\url{https://doi.org/10.5281/zenodo.15576593}).}


\conflict{The authors declare they have no conflict of interests.}

\aucontribute{J.G.T.R. and A.C. jointly conceived the work plan. J.G.T.R. developed the mathematical model and conducted numerical simulations, while A.C. reviewed the mathematical development and provided insights into the mechanical interpretations of the model. Both authors discussed the results extensively and contributed to the writing of the manuscript. J.G.T.R. and A.C. reviewed and approved the final version of the manuscript.}

\funding{The authors acknowledge support from the National Council for Scientific and Technological Development (CNPq), grant 305476/2022-0; the Carlos Chagas Filho Research Foundation of Rio de Janeiro State (FAPERJ) grant 204.477/2024; and the National Institute of Science and Technology - Smart Structures in Engineering (INCT-EIE), funded by CNPq under grant number 406148/2022-8, as well as Coordination for the Improvement of Higher Education Personnel (CAPES) and Minas Gerais State Research Support Foundation (FAPEMIG).}

\ack{The first author acknowledges the useful discussion about creeping behavior with Prof. Jaime Tupiassú Pinho de Castro and Prof. Marco Antonio Meggiolaro, both from PUC-Rio (Brazil). The useful comments and critics from the anonymous reviewers are also very appreciated, since they contributed a lot to improve the final version of this paper.}

\disclaimer{The information provided in this paper as well as the interpretation of the results is solely that by the authors, and it does not necessarily reflect the views of the sponsors or sponsoring agencies.}

\def\ai#1{{\vskip5.5pt\noindent \textcolor{jobcolor}{\fontsize{9}{11}\selectfont Declaration of AI use.}\fontsize{8}{11}\selectfont\enskip #1}}

\ai{The authors declare that AI tools, including ChatGPT and Grammarly, were utilized to enhance text clarity and writing quality; however, the authors retain full responsibility for the wording and organization of the paper.}

\vskip2pc

\bibliographystyle{elsarticle-num}

\begin{thebibliography}{10}
\expandafter\ifx\csname url\endcsname\relax
  \def\url#1{\texttt{#1}}\fi
\expandafter\ifx\csname urlprefix\endcsname\relax\def\urlprefix{URL }\fi
\expandafter\ifx\csname href\endcsname\relax
  \def\href#1#2{#2} \def\path#1{#1}\fi

\bibitem{crawford2020plastics}
R.~J. Crawford, P.~Martin, Plastics Engineering, Butterworth-Heinemann, Oxford,
  United Kingdom, 2020.

\bibitem{Dowling2012}
N.~E. Dowling, Mechanical Behavior of Materials, 4th Edition, Prentice Hall,
  2012.

\bibitem{castro2016fatigue}
J.~T.~P. Castro, M.~A. Meggiolaro, Fatigue Design Techniques, Volume 3: Crack
  Propagation, Temperature and Statistical Effects, CreateSpace, South
  Carolina, US, 2016.

\bibitem{morro2017modelling}
A.~Morro, Modelling of viscoelastic materials and creep behaviour, Meccanica 52
  (2017) 3015--3021.
\newblock \href {https://doi.org/10.1007/s11012-016-0585-x}
  {\path{doi:10.1007/s11012-016-0585-x}}.

\bibitem{Lakes2010}
R.~Lakes, Viscoelastic Materials, Cambridge University Press, Cambridge, UK,
  2010.

\bibitem{di2011visco}
M.~Di~Paola, A.~Pirrotta, A.~Valenza, Visco-elastic behavior through fractional
  calculus: {A}n easier method for best fitting experimental results, Mechanics
  of Materials 43 (2011) 799--806.
\newblock \href {https://doi.org/10.1016/j.mechmat.2011.08.016}
  {\path{doi:10.1016/j.mechmat.2011.08.016}}.

\bibitem{nutting1921new}
P.~Nutting, A new general law of deformation, Journal of the Franklin Institute
  191 (1921) 679--685.
\newblock \href {https://doi.org/10.1016/S0016-0032(21)90171-6}
  {\path{doi:10.1016/S0016-0032(21)90171-6}}.

\bibitem{gemant1936method}
A.~Gemant, A method of analyzing experimental results obtained from
  elasto-viscous bodies, Physics 7 (1936) 311--317.
\newblock \href {https://doi.org/10.1063/1.1745400}
  {\path{doi:10.1063/1.1745400}}.

\bibitem{mainardi2008time}
F.~Mainardi, R.~Gorenflo, Time-fractional derivatives in relaxation processes:
  {A} tutorial survey, Fractional Calculus and Applied Analysis 10 (2007)
  269--308.

\bibitem{blair1944analytical}
G.~S. Blair, Analytical and integrative aspects of the stress-strain-time
  problem, Journal of Scientific Instruments 21 (1944) 80.
\newblock \href {https://doi.org/10.1088/0950-7671/21/5/302}
  {\path{doi:10.1088/0950-7671/21/5/302}}.

\bibitem{ortigueira2015fractional}
M.~D. Ortigueira, J.~A. {Tenreiro Machado}, What is a fractional derivative?,
  Journal of Computational Physics 293 (2015) 4--13.
\newblock \href {https://doi.org/10.1016/j.jcp.2014.08.016}
  {\path{doi:10.1016/j.jcp.2014.08.016}}.

\bibitem{ribeiro2021modeling}
J.~G.~T. Ribeiro, J.~T.~P. de~Castro, M.~A. Meggiolaro, Modeling concrete and
  polymer creep using fractional calculus, Journal of Materials Research and
  Technology 12 (2021) 1184--1193.
\newblock \href {https://doi.org/10.1016/j.jmrt.2021.03.007}
  {\path{doi:10.1016/j.jmrt.2021.03.007}}.

\bibitem{bonfanti2020fractional}
A.~Bonfanti, J.~L. Kaplan, G.~Charras, A.~Kabla, Fractional viscoelastic models
  for power-law materials, Soft Matter 16 (2020) 6002--6020.
\newblock \href {https://doi.org/10.1039/D0SM00354A}
  {\path{doi:10.1039/D0SM00354A}}.

\bibitem{jozwiak2015fractional}
B.~J{\'o}{\'z}wiak, M.~Orczykowska, M.~Dziubi{\'n}ski, Fractional
  generalizations of maxwell and kelvin-voigt models for biopolymer
  characterization, PloS one 10 (2015) e0143090.
\newblock \href {https://doi.org/10.1371/journal.pone.0143090}
  {\path{doi:10.1371/journal.pone.0143090}}.

\bibitem{di2020novel}
M.~Di~Paola, G.~Alotta, A.~Burlon, G.~Failla, A novel approach to nonlinear
  variable-order fractional viscoelasticity, Philosophical Transactions of the
  Royal Society A 378 (2020) 20190296.
\newblock \href {https://doi.org/10.1098/rsta.2019.0296}
  {\path{doi:10.1098/rsta.2019.0296}}.

\bibitem{su2020fractional}
X.~Su, W.~Xu, W.~Chen, H.~Yang, Fractional creep and relaxation models of
  viscoelastic materials via a non-{N}ewtonian time-varying viscosity:
  {P}hysical interpretation, Mechanics of Materials 140 (2020) 103222.
\newblock \href {https://doi.org/10.1016/j.mechmat.2019.103222}
  {\path{doi:10.1016/j.mechmat.2019.103222}}.

\bibitem{gao2023bridge}
Y.~Gao, D.~Yin, B.~Zhao, A bridge between the fractional viscoelasticity and
  time-varying viscosity model: {P}hysical interpretation and constitutive
  modeling, Mechanics of Time-Dependent Materials 27 (2023) 1153--1170.
\newblock \href {https://doi.org/10.1007/s11043-022-09555-y}
  {\path{doi:10.1007/s11043-022-09555-y}}.

\bibitem{garrappa2021variable}
R.~Garrappa, A.~Giusti, F.~Mainardi, Variable-order fractional calculus: {A}
  change of perspective, Communications in Nonlinear Science and Numerical
  Simulation 102 (2021) 105904.
\newblock \href {https://doi.org/10.1016/j.cnsns.2021.105904}
  {\path{doi:10.1016/j.cnsns.2021.105904}}.

\bibitem{patnaik2020applications}
S.~Patnaik, J.~P. Hollkamp, F.~Semperlotti, Applications of variable-order
  fractional operators: {A} review, Proceedings of the Royal Society A 476
  (2020) 20190498.
\newblock \href {https://doi.org/10.1098/rspa.2019.0498}
  {\path{doi:10.1098/rspa.2019.0498}}.

\bibitem{samko1993integration}
S.~G. Samko, B.~Ross, Integration and differentiation to a variable fractional
  order, Integral Transforms and Special Functions 1 (1993) 277--300.
\newblock \href {https://doi.org/10.1080/10652469308819027}
  {\path{doi:10.1080/10652469308819027}}.

\bibitem{samko1995fractional}
S.~G. Samko, Fractional integration and differentiation of variable order,
  Analysis Mathematica 21 (1995) 213--236.
\newblock \href {https://doi.org/10.1007/bf01911126}
  {\path{doi:10.1007/bf01911126}}.

\bibitem{samko2013fractional}
S.~Samko, Fractional integration and differentiation of variable order: an
  overview, Nonlinear Dynamics 71 (2013) 653--662.
\newblock \href {https://doi.org/10.1007/s11071-012-0485-0}
  {\path{doi:10.1007/s11071-012-0485-0}}.

\bibitem{ramirez2007variable}
L.~E. Ramirez, C.~F. Coimbra, A variable order constitutive relation for
  viscoelasticity, Annalen der Physik 519 (2007) 543--552.
\newblock \href {https://doi.org/10.1002/andp.200710246}
  {\path{doi:10.1002/andp.200710246}}.

\bibitem{cai2020variable}
W.~Cai, P.~Wang, J.~Fan, A variable-order fractional model of tensile and shear
  behaviors for sintered nano-silver paste used in high power electronics,
  Mechanics of Materials 145 (2020) 103391.
\newblock \href {https://doi.org/10.1016/j.mechmat.2020.103391}
  {\path{doi:10.1016/j.mechmat.2020.103391}}.

\bibitem{gao2021full}
Y.~Gao, D.~Yin, A full-stage creep model for rocks based on the variable-order
  fractional calculus, Applied Mathematical Modelling 95 (2021) 435--446.
\newblock \href {https://doi.org/10.1016/j.apm.2021.02.020}
  {\path{doi:10.1016/j.apm.2021.02.020}}.

\bibitem{wang2023fractional}
P.~Wang, W.~Cai, Y.~Zhang, Z.~Wang, A fractional rheological model for
  loading-dependent rheological behavior of polymers, Mechanics of
  Time-Dependent Materials (2023) 1--12\href
  {https://doi.org/10.1007/s11043-023-09616-w}
  {\path{doi:10.1007/s11043-023-09616-w}}.

\bibitem{li2022variable}
Z.~Li, Z.~Dong, Z.~Zhang, B.~Han, B.~Sun, Y.~Wang, F.~Liu, Variable-order
  fractional dynamic behavior of viscoelastic damping material, Journal of
  Mechanics 38 (2022) 323--332.
\newblock \href {https://doi.org/10.1093/jom/ufac025}
  {\path{doi:10.1093/jom/ufac025}}.

\bibitem{meng2019variable}
R.~Meng, D.~Yin, C.~S. Drapaca, A variable order fractional constitutive model
  of the viscoelastic behavior of polymers, International Journal of Non-Linear
  Mechanics 113 (2019) 171--177.
\newblock \href {https://doi.org/10.1016/j.ijnonlinmec.2019.04.002}
  {\path{doi:10.1016/j.ijnonlinmec.2019.04.002}}.

\bibitem{meng2020parameter}
R.~Meng, D.~Yin, H.~Yang, G.~Xiang, Parameter study of variable order
  fractional model for the strain hardening behavior of glassy polymers,
  Physica A: Statistical Mechanics and its Applications 545 (2020) 123763.
\newblock \href {https://doi.org/10.1016/j.physa.2019.123763}
  {\path{doi:10.1016/j.physa.2019.123763}}.

\bibitem{xiang2020creep}
G.~Xiang, D.~Yin, R.~Meng, S.~Lu, Creep model for natural fiber polymer
  composites (nfpcs) based on variable order fractional derivatives: Simulation
  and parameter study, Journal of Applied Polymer Science 137 (2020) 48796.
\newblock \href {https://doi.org/10.1002/app.48796}
  {\path{doi:10.1002/app.48796}}.

\bibitem{patnaik2020variable}
S.~Patnaik, F.~Semperlotti, Variable-order particle dynamics: {F}ormulation and
  application to the simulation of edge dislocations, Philosophical
  Transactions of the Royal Society A 378 (2020) 20190290.
\newblock \href {https://doi.org/10.1098/rsta.2019.0290}
  {\path{doi:10.1098/rsta.2019.0290}}.

\bibitem{patnaik2021variable}
S.~Patnaik, F.~Semperlotti, Variable-order fracture mechanics and its
  application to dynamic fracture, npj Computational Materials 7 (2021) 27.
\newblock \href {https://doi.org/10.1038/s41524-021-00492-x}
  {\path{doi:10.1038/s41524-021-00492-x}}.

\bibitem{beltempo2018fractional}
A.~Beltempo, M.~Zingales, O.~S. Bursi, L.~Deseri, A fractional-order model for
  aging materials: An application to concrete, International Journal of Solids
  and Structures 138 (2018) 13--23.
\newblock \href {https://doi.org/10.1016/j.ijsolstr.2017.12.024}
  {\path{doi:10.1016/j.ijsolstr.2017.12.024}}.

\bibitem{machado2011recent}
J.~A. {Tenreiro Machado}, V.~Kiryakova, F.~Mainardi, Recent history of
  fractional calculus, Communications in Nonlinear Science and Numerical
  Simulation 16~(3) (2011) 1140--1153.
\newblock \href {https://doi.org/10.1016/j.cnsns.2010.05.027}
  {\path{doi:10.1016/j.cnsns.2010.05.027}}.

\bibitem{machado2014development}
J.~A. {Tenreiro Machado}, A.~Galhano, J.~Trujillo, On development of fractional
  calculus during the last fifty years, Scientometrics 98 (2014) 577--582.
\newblock \href {https://doi.org/10.1007/s11192-013-1062-7}
  {\path{doi:10.1007/s11192-013-1062-7}}.

\bibitem{delia2021unified}
M.~D’Elia, M.~Gulian, H.~Olson, et~al., Towards a unified theory of
  fractional and nonlocal vector calculus, Fractional Calculus and Applied
  Analysis 24 (2021) 1301--1355.
\newblock \href {https://doi.org/10.1515/fca-2021-0057}
  {\path{doi:10.1515/fca-2021-0057}}.

\bibitem{colombaro2018scott}
I.~Colombaro, R.~Garra, A.~Giusti, F.~Mainardi, {Scott-Blair} models with
  time-varying viscosity, Applied Mathematics Letters 86 (2018) 57--63.
\newblock \href {https://doi.org/10.1016/j.aml.2018.06.022}
  {\path{doi:10.1016/j.aml.2018.06.022}}.

\bibitem{almeida2015computing}
R.~Almeida, D.~F. Torres, Computing {H}adamard type operators of variable
  fractional order, Applied Mathematics and Computation 257 (2015) 74--88.

\bibitem{CunhaJr2024CEopt}
A.~{Cunha~Jr}, M.~V. Issa, J.~C. Basilio, J.~G. {Telles Ribeiro}, {CEopt: A
  MATLAB Package for Non-convex Optimization with the Cross-Entropy Method},
  arXiv:2409.00013 (2024).
\newblock \href {https://doi.org/10.48550/arXiv.2409.00013}
  {\path{doi:10.48550/arXiv.2409.00013}}.

\bibitem{kroese2013handbook}
D.~P. Kroese, T.~Taimre, Z.~I. Botev, Handbook of monte carlo methods, John
  Wiley \& Sons, Hoboken, NJ, USA, 2013.

\bibitem{cunha2014uncertainty}
A.~Cunha~Jr, R.~Nasser, R.~Sampaio, H.~Lopes, K.~Breitman, Uncertainty
  quantification through the monte carlo method in a cloud computing setting,
  Computer Physics Communications 185 (2014) 1355--1363.
\newblock \href {https://doi.org/10.1016/j.cpc.2014.01.006}
  {\path{doi:10.1016/j.cpc.2014.01.006}}.

\bibitem{cunha2021enhancing}
A.~Cunha~Jr, Enhancing the performance of a bistable energy harvesting device
  via the cross-entropy method, Nonlinear Dynamics 103 (2021) 137--155.
\newblock \href {https://doi.org/10.1007/s11071-020-06109-0}
  {\path{doi:10.1007/s11071-020-06109-0}}.

\bibitem{Issa2023cobem}
M.~V.~S. Issa, A.~{Cunha~Jr}, F.~J. C.~P. Soeiro, A.~Pereira, Optimizing truss
  structures with natural frequency constraints using the cross-entropy method,
  in: 27th International Congress of Mechanical Engineering (COBEM~2023),
  Florian\'opolis, Brazil, 2023.

\bibitem{Dantas2019cobem}
E.~Dantas, A.~{Cunha~Jr}, F.~J. C.~P. Soeiro, B.~C. Cayres, H.~I. Weber, An
  inverse problem via cross-entropy method for calibration of a drill string
  torsional dynamic model, in: 25th ABCM International Congress of Mechanical
  Engineering (COBEM~2019), Uberl\^{a}ndia, Brazil, 2019.

\bibitem{Dantas2019icedyn}
E.~Dantas, A.~{Cunha~Jr}, T.~A.~N. Silva, A numerical procedure based on
  cross-entropy method for stiffness identification, in: 5th International
  Conference on Structural Engineering Dynamics (ICEDyn~2019), Viana do
  Castelo, Portugal, 2019.

\end{thebibliography}


\end{document}